\documentclass[fleqn,usenatbib]{mnras}

\usepackage{newtxtext,newtxmath}

\usepackage[T1]{fontenc}

\DeclareRobustCommand{\VAN}[3]{#2}
\let\VANthebibliography\thebibliography
\def\thebibliography{\DeclareRobustCommand{\VAN}[3]{##3}\VANthebibliography}

\usepackage{graphicx}	
\usepackage{amsmath}	

\title[Radio Emission Regions in MSPs]{Radio emission from beyond the light cylinder in millisecond pulsars}

\author[Kramer \& Johnston]{
Michael Kramer,$^{1,2}$\thanks{E-mail: mkramer@mpifr-bonn.mpg.de}
Simon Johnston$^{3}$
\\
$^1$Max-Planck-Institut f{\"u}r Radioastronomie, Auf dem H{\"u}gel 69, D-53121 Bonn, Germany\\
$^2$Jodrell Bank Centre for Astrophysics, University of Manchester, Manchester M13 9PL, UK \\
$^{3}$Australia Telescope National Facility, CSIRO, Space and Astronomy, PO Box 76, Epping, NSW 1710, Australia\\
}

\date{Accepted XXX. Received YYY; in original form ZZZ}

\pubyear{2025}

\begin{document}
\label{firstpage}
\pagerange{\pageref{firstpage}--\pageref{lastpage}}
\maketitle

\begin{abstract}
A striking aspect of the radio profiles of many millisecond pulsars (MSPs) is that they consist of components separated from each other by regions lacking in emission. We devise a technique for determining ``disjoint" from ``contiguous" components and show that 35\% of MSPs have disjoint components as opposed to only 3\% of the slow pulsar population. We surmise that the pulsars with these disjoint components show evidence for both emission above the polar cap and from the current sheet beyond the light cylinder (LC), co-located with gamma-ray emission. For some of the radio MSPs only the LC emission is being observed. It is our contention that almost all of the current population of gamma-ray MSPs show evidence for co-located radio emission. A simple geometric explanation allows the presence (or not) of LC emission and the relationship (or not) between the gamma-ray and radio profiles to be determined. The LC components have frequently very high polarization and typically flat position-angle traverses thus helping to explain the difficulties in determining the geometry of MSPs. In cases where the geometry can be determined the values broadly align with expectations. In this picture, the number of potentially detectable radio MSPs is higher than previously thought, although the actual detectability of LC components depends on their luminosity function. A mechanism is required to produce coherent radio emission far from the stellar surface. 
These ideas have implications for our understanding of the populations of radio-loud and radio-quiet rotation-powered millisecond pulsars, and may have implications for the long-term timing stability of some of these sources.
\end{abstract}




\section{Introduction}
A common tool for visualising the pulsar population is the $P$-$\dot{P}$ plane with $P$ the pulsar's spin period and $\dot{P}$ its rate of slow-down. Broadly speaking, the $P$-$\dot{P}$ diagram shows two distinct populations. The first and larger group is the `slow' pulsars; these pulsars are predominantly single, have $P>100$~ms, high magnetic fields and ages around 1~Myr. In contrast, the recycled or millisecond pulsars (MSPs) are often part of binary systems, have $P<50$~ms, low magnetic fields and ages in excess of 1~Gyr. In spite of these differences, many of the emission properties of the two classes are remarkably similar \citep{paperI,paperII,paperIII}. As summarised in a recent paper by \citet{kjp+24}, the MSPs and slow pulsars have similar spectra, profiles and luminosities and the widths of the MSP profiles obey the same period relationship as the slow pulsars. These similarities are largely true for the polarisation properties, although MSPs more frequently show extremely flat position angle swings \citep{paperII,relbin}.

However, there appears to be a striking visual difference between the radio profiles of slow pulsars and MSPs. In slow pulsars, the profiles are made up of components which, in the vast majority of cases, occupy a contiguous region of pulse longitude (see e.g. the collections of \citealt{mbm+16}, \citealt{jk18}, \citealt{pkj+23} and \citealt{whx+23}). The exceptions are generally identified as interpulse pulsars. Here, two distinct regions of pulse longitude are occupied separated by 180\degr; emission from both poles is being seen \citep{jk19,swyw25}. In contrast, the millisecond pulsars often show distinct regions of emission with zero emission between the regions, as shown by the profiles given in e.g. \citet{paperI}, \citet{dai15}, \citet{wmg+22}, \citet{sbm+22}
or \citet{xjx+25}.

Traditionally, it has been thought that the radio emission arises from near the surface of the star, in a region bounded by the last open field-lines. The emission height for slow pulsars appears to be some 300~km, largely independent of spin period \citep{pkj+23}. This height is at most only a few percent of the light-cylinder radius. For the faster rotating millisecond pulsars the light cylinder radius is only of the order of 100~km and so (in this picture) the radio emission must occur from only a few stellar radii.

Since the launch of the Fermi satellite, a large number of both slow and millisecond pulsars are now known to emit gamma-rays \citep{3PC}. The location of the gamma-ray emission also remains an open question, but prevalent models call for the emission to arise from either just inside the light cylinder \citep{har16} or in the equatorial current sheet outside the light cylinder \citep{ksg02,bs10,bpm21,kwhk23,pet24,cfd25}. Several of the MSPs have gamma-ray profiles that align with the radio profiles with the implication that the radio and gamma-ray emission regions are physically co-located \citep{Abdo0034,vjh12,gjv+12}. In the young objects PSRs~J0540--6919 and the Crab, the giant pulse emission in the radio is in phase with the high-energy emission and distinct from the non-giant radio emission \citep{jr04,lcu+95}. However, in the rest of the slow pulsars, the gamma-ray and radio profiles are misaligned.
\begin{figure*}
\begin{center}
\begin{tabular}{cc}
\includegraphics[width=8cm]{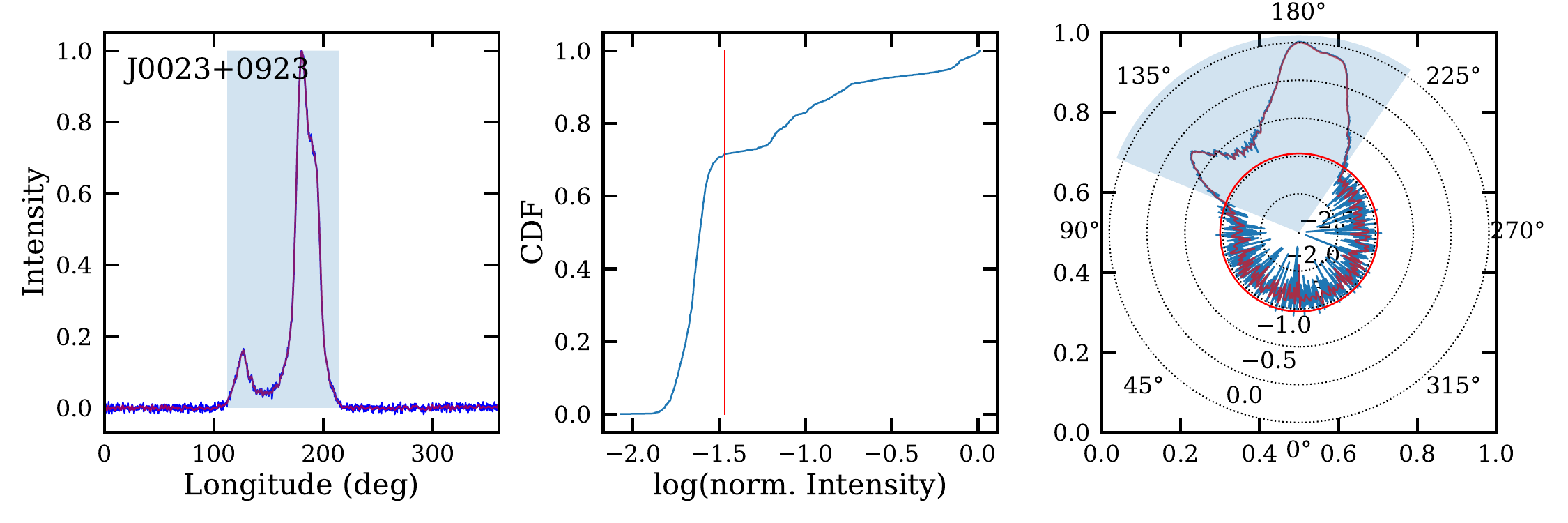}  &
\includegraphics[width=8cm]{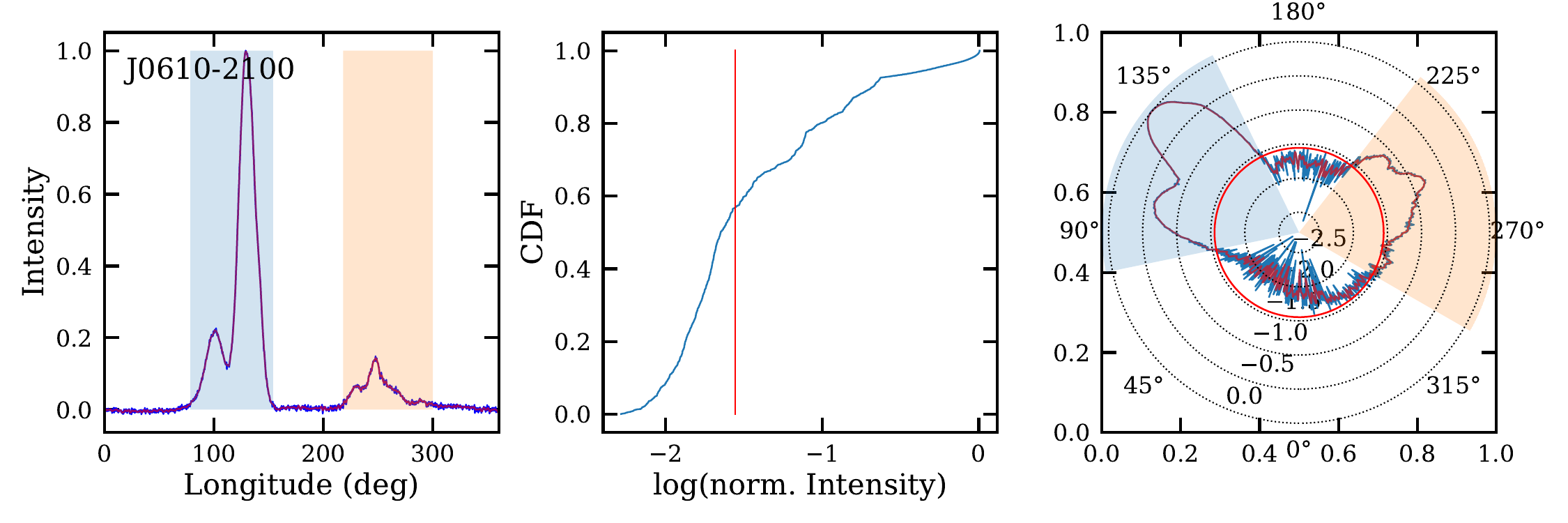}  \\
\includegraphics[width=8cm]{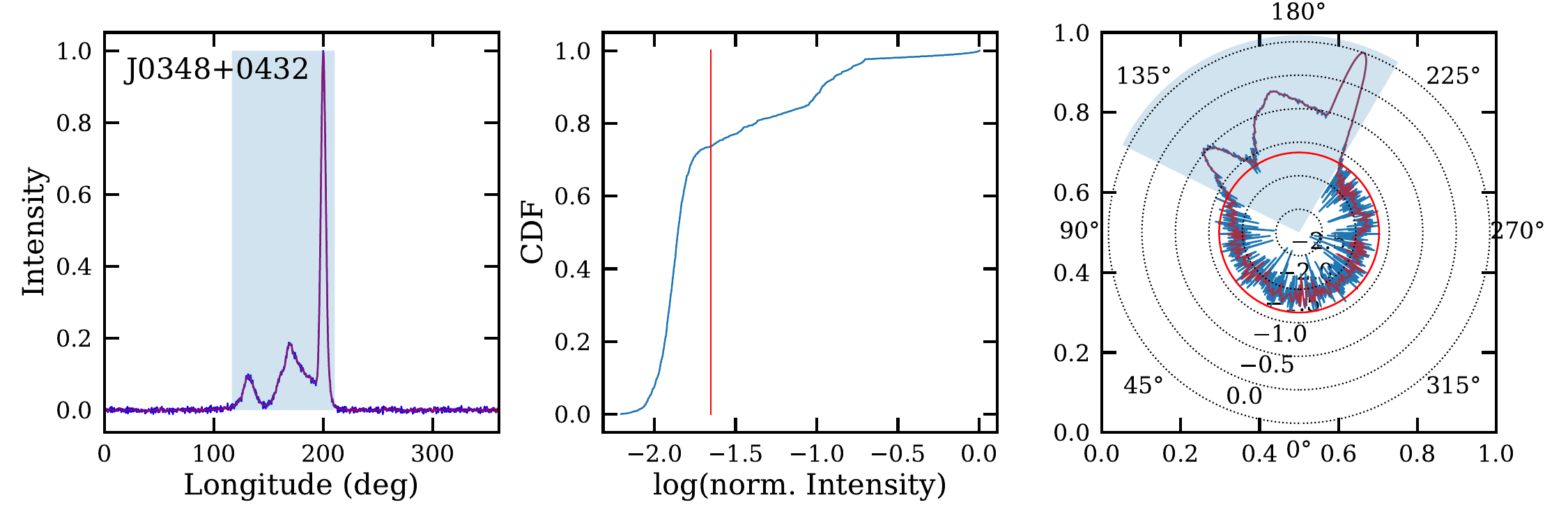}  &
\includegraphics[width=8cm]{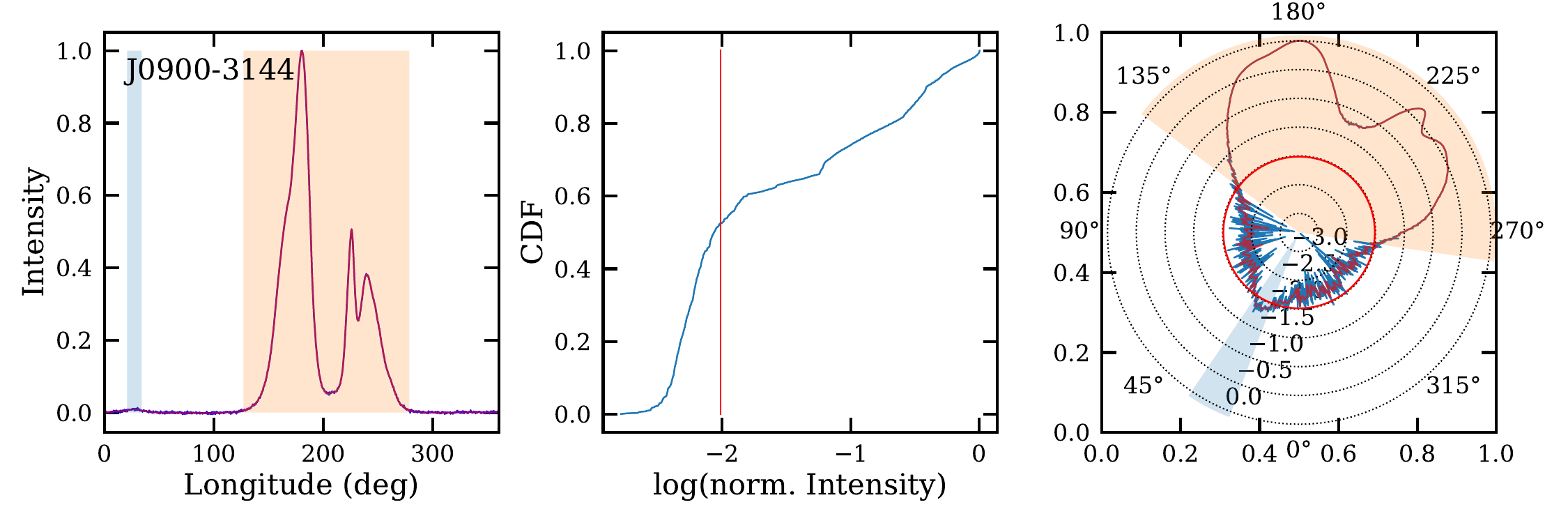}  \\
\includegraphics[width=8cm]{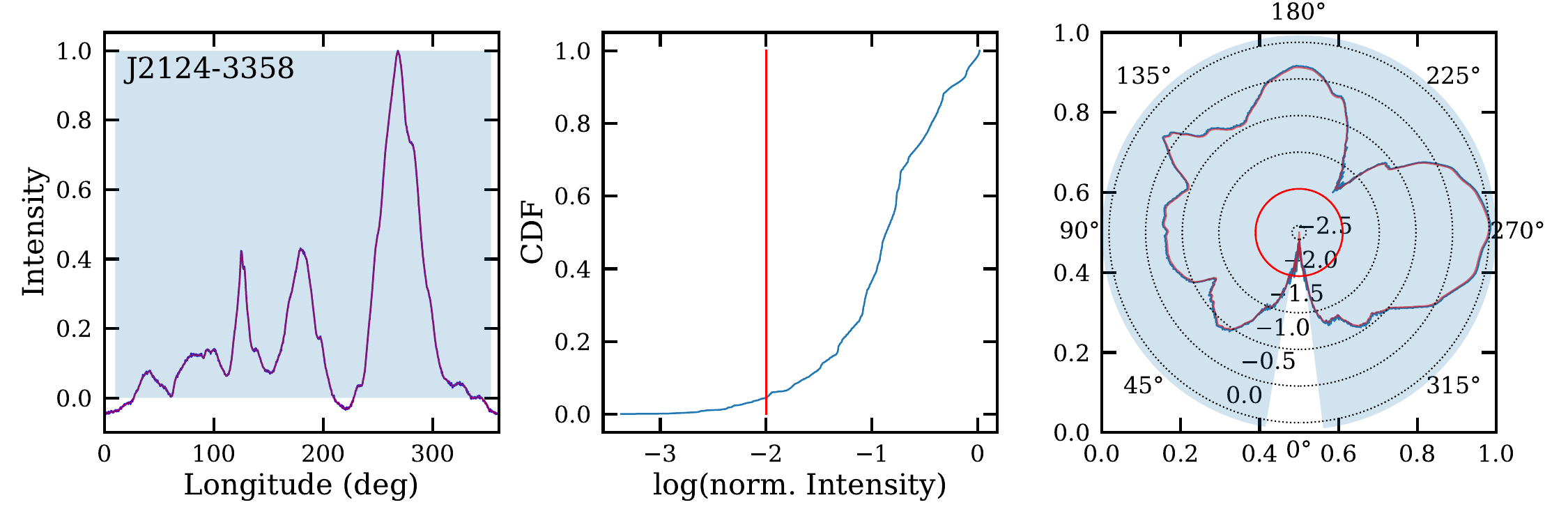} &
\includegraphics[width=8cm]{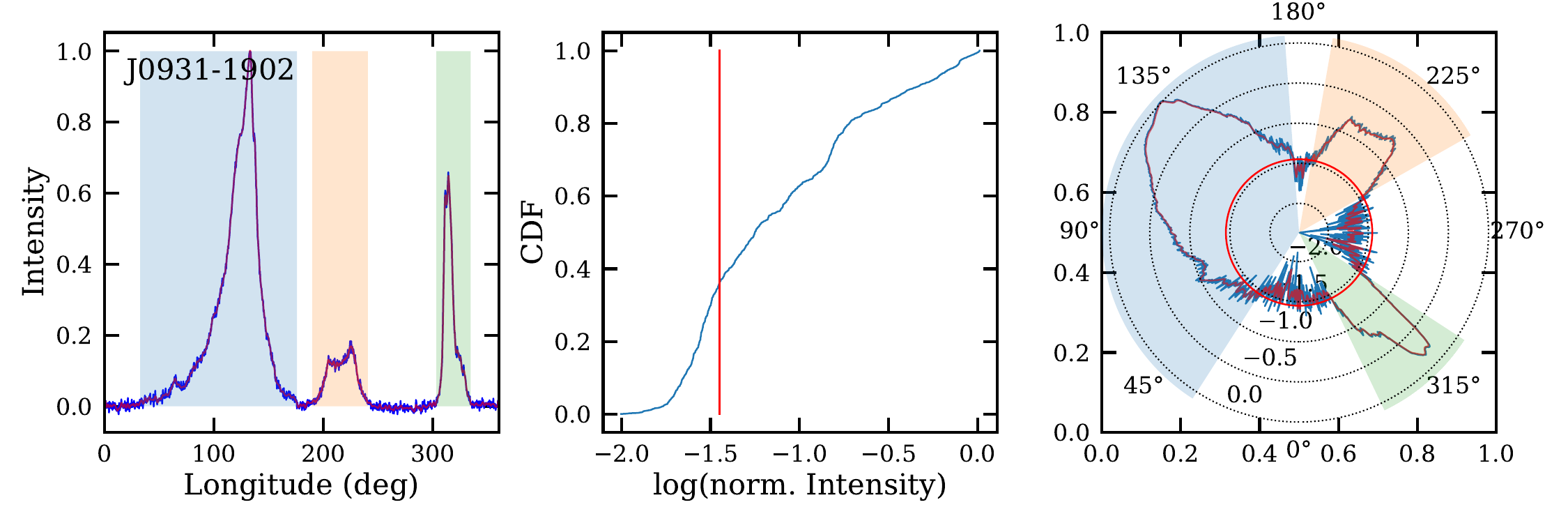} \\
\end{tabular}
\end{center}
\caption{Example outputs from the finding technique for six millisecond pulsars. The left hand column shows pulsars classified as contiguous (class C) and the right hand column shows pulsars with disjoint (class D) profiles. The three panels show the profile, the cumulative distribution of the amplitudes and the polar plot. See text for details.}
\label{fig:msp}
\end{figure*}

The visibility (or not) of emission from beyond the light cylinder in the equatorial current sheet is strongly confined by the geometry of the star and the viewing angle (e.g. \citealt{pet24}). Similarly the visibility (or not) of the polar cap is also geometry dependent. Hence the detection of radio and/or gamma-ray emission from a given pulsar can be used to constrain its geometry (e.g. \citealt{jskk20,bpm21}).

In this paper we re-examine the radio emission regions for the millisecond pulsars using the excellent observational datasets now at our disposal. We especially acknowledge the work of \citet{3PC} who provide phase-aligned gamma-ray and radio profiles. In Section~2 we outline the technique used to distinguish different profile types. Section~3 present the results for the slow and millisecond pulsar populations, and we discuss the implications of these results in Section~4.

\section{Technique}
\subsection{Profiles}
The emission features identified at pulse phases distinct from the main pulse or interpulse often appear at very low intensity level relative to the other parts of the profile \citep{paperI}. In order to detect and characterise these regions of pulse phase, we investigated the noise properties of the off-pulse regions and use log-polar plots, first introduced (to our knowledge) by \cite{hf86}. Examples of these plots are shown in Figure~\ref{fig:msp}. Three panels are shown for each pulsar. The left panel shows the profile, normalised by the peak. The coloured regions identify contiguous regions of emission (each marked with a separate colour) and are determined as follows. 

For each profile, we subtract the minimum of the off-pulse region as an offset to avoid negative numbers. We derive a smoothed version of this profile by deploying a Gaussian Process with a RBF kernel using the {\tt GEORGE}\footnote{\tt https://github.com/dfm/george} package \citep{george}. We sort the logarithm of this profile version and compute the cumulative distribution shown in the middle panel. Examining this curve, we can usually identify a break-point  that separates the noisy off-pulse regions from those with significant radio emission. In most cases, this break-point coincides with 5 times the RMS value measured in the off-pulse regions of the original data, indicated by the red vertical line. The original profile (blue) and its smoothed version (red) are shown in the left panel and in the polar plot in the right panel. The intensity level identified by the red vertical line from the cumulative distributions (middle panel) is shown as a red circle in the right panel. We then determine the regions where the emission exceeds the red intensity level. Regions separated by less than two degrees of longitude (or 6 bins for a 1024-bin profile) are merged and considered as one region. Similarly, regions that are less than two degrees wide are ignored. Finally, we record the centre and width of these regions and measure their maximum intensity relative to the peak.

For some pulsars that show emission across nearly 360\degr\ of longitude, an off-pulse region, its RMS value and, correspondingly, a break-point are difficult to measure (see e.g.~PSR J2124$-$3358 in Figure~\ref{fig:msp}). In these cases, we adjusted the red intensity level by eye, guided by the polar plot, where the contiguous regions are often easy to identify.

Based on this method, we classify the pulsars into two categories. Pulsars with a single contiguous region of emission are denoted as Class C whereas pulsars with two or more contiguous emission regions separated by regions below the noise level are Class D.
\begin{figure}
\begin{center}
\begin{tabular}{c}
\includegraphics[width=8cm]{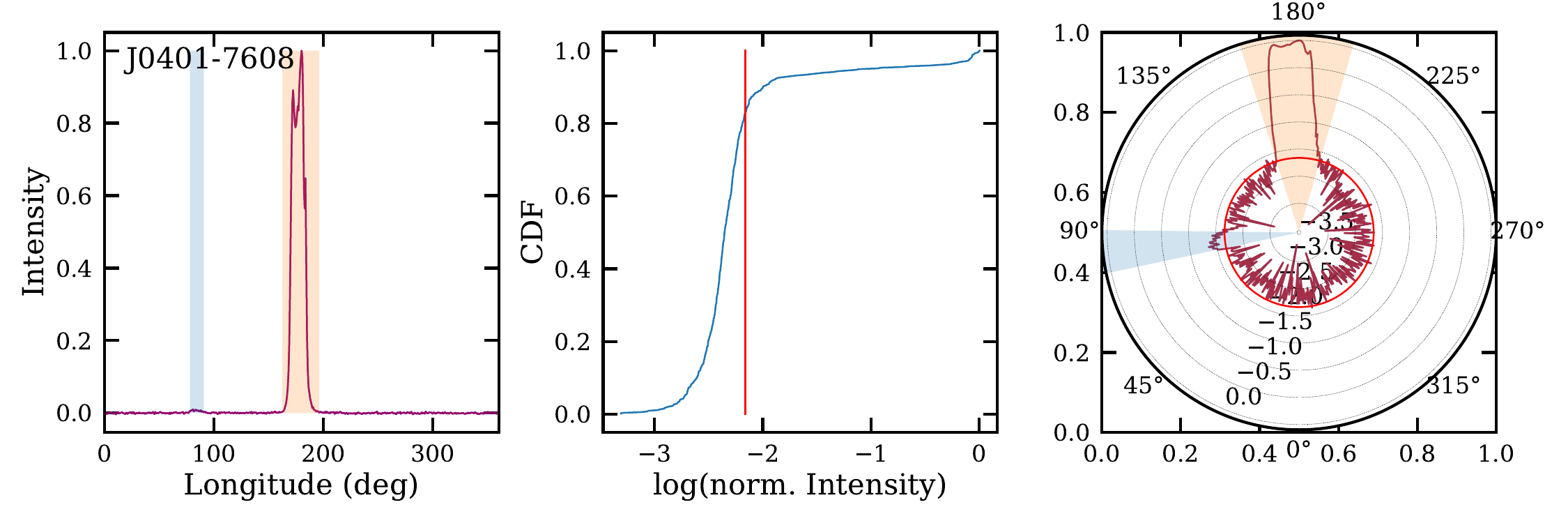}  \\
\includegraphics[width=8cm]{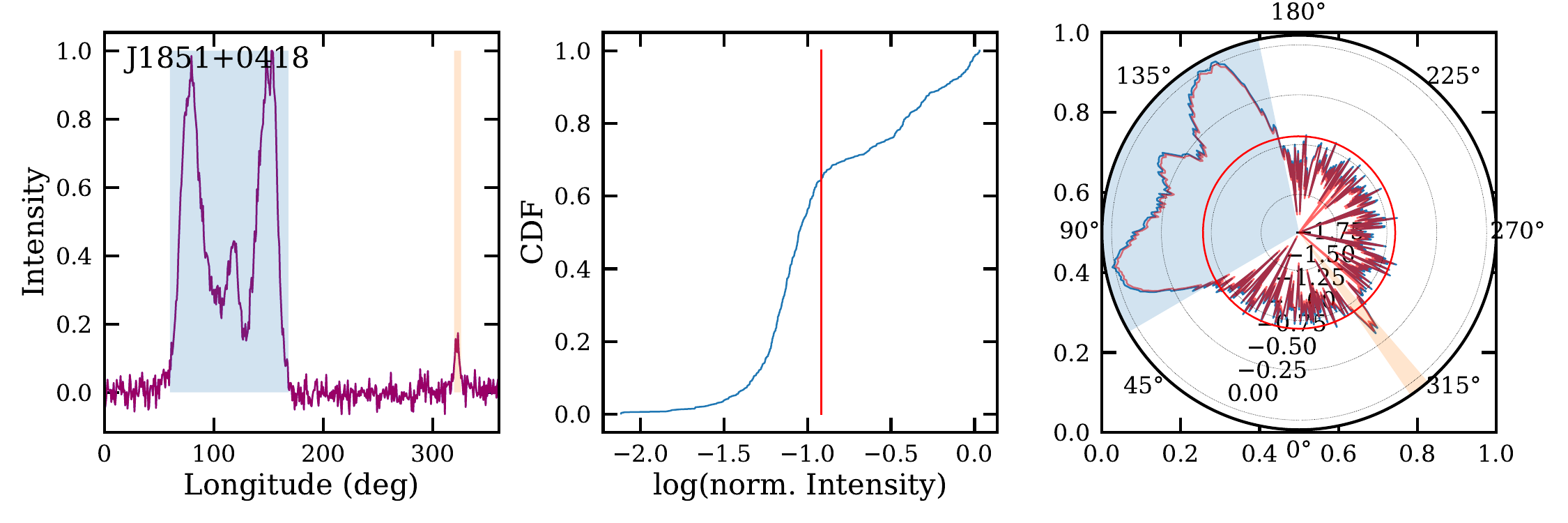}  \\
\includegraphics[width=8cm]{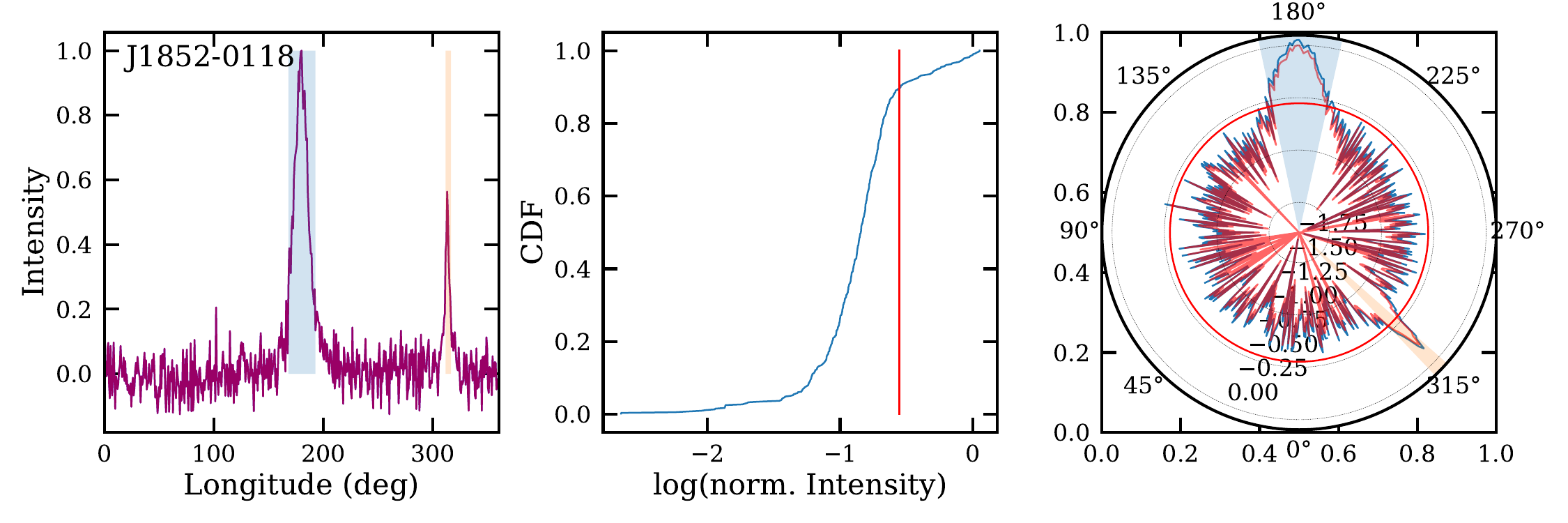}  \\
\end{tabular}
\end{center}
\caption{Example outputs from the finding technique for three slow pulsars with disjoint profiles, PSRs J0401--7608 (top), J1851+0418 (middle) and J1852--0118 (bottom). See Figure~\ref{fig:msp} and text for details.}
\label{fig:slow}
\end{figure}

\subsection{Geometry}
In the rotating vector model (RVM) of \citet{rc69a}, the radiation is beamed along the field lines and the plane of polarization is determined by the angle of the magnetic field as it sweeps past the line of sight. The position angle (PA) as a function of pulse longitude, $\phi$, can be expressed as
\begin{equation}
\label{eqn:rvm}
{\rm PA} = {\rm PA}_{0} +
{\rm arctan} \left( \frac{{\rm sin}\alpha
\, {\rm sin}(\phi - \phi_0)}{{\rm sin}\zeta
\, {\rm cos}\alpha - {\rm cos}\zeta
\, {\rm sin}\alpha \, {\rm cos}(\phi - \phi_0)} \right)
\end{equation}
Here, $\alpha$ is the angle between the rotation axis and the magnetic axis and $\zeta=\alpha+\beta$ with $\beta$ being the angle of closest approach of the line of sight to the magnetic axis. $\phi_0$ is the longitude at which the PA is PA$_{0}$, which corresponds to the PA of the rotation axis projected onto the plane of the sky. We note that RVM fitting does not depend on the total intensity profile, or the location of the profile symmetry points.

The angles in Eq.~\ref{eqn:rvm} are defined as in \cite{rc69a}, i.e.\ using the `RVM/DT92' convention \citep[see e.g.][]{ew01,relbin}.  We determine the RVM parameter posteriors in their joint parameter space following the method outlined in \cite{jk19} and \cite{relbin}. The method automatically accounts for the possibility of `orthogonal jumps', where the PA transitions from the RVM track to one 90\degr\ apart (e.g. \citealt{lk05}). Quoted geometries and joint parameter uncertainties are determined from corner plots of the posterior distributions. 

\section{Results}
\subsection{MSPs}
For the millisecond pulsar sample we used the profiles given in \citet{sbm+22}, \citet{whx+23}, \citet{xjx+25} and \citet{wmg+22}. We adopt an upper limit of 25~ms on the spin period and excluded a small number of other pulsars with low s/n and/or scatter-broadened profiles.  This yielded a total of 194 unique pulsars, none of which are in globular clusters. Of these, 119 pulsars (61\%) are Class C and 75 pulsars (39\%) show disjoint emission, Class D. Figure~\ref{fig:msp} shows the output for a sample of six millisecond pulsars, three each from Classes C and D. The polar plots (right most panels) are particularly striking and allow for easy visualisation of the various components.

We note that the profiles vary significantly in s/n from one pulsar to the next. Some, such as PSR~J0437$-$4715 has a peak flux some $10^4$ times higher than the baseline level whereas in others such as PSR~J0101$-$6422 this ratio is only 10. Low-level components or weak ``bridge" emission between components may be missed in these lower s/n profiles. Where multiple groups have observed the same pulsar we use the profile with the highest s/n.

Of the 194 pulsars, 81 are gamma-ray emitters with profiles available in \citet{3PC}, a further 85 have been searched for gamma-rays without success and 33 do not have ephemerides good enough to perform gamma-ray searches. A substantially higher fraction (69\%) of Class D pulsars are gamma-ray detected compared to Class C (24\%).

\subsection{Slow Pulsars}
For the slow pulsars we used the profiles from \citet{pkj+23} and \citet{whx+23}. From a total of 1611 unique pulsars we deemed 1492 suitable for analysis. Of these, only 51 pulsars (3.4\%) fall into Class D. These pulsars will be discussed in more detail in Section 4.2. A total of 55 of the slow pulsars have been detected in gamma-rays \citep{3PC}.

\begin{figure}
\begin{center}
\includegraphics[width=8cm]{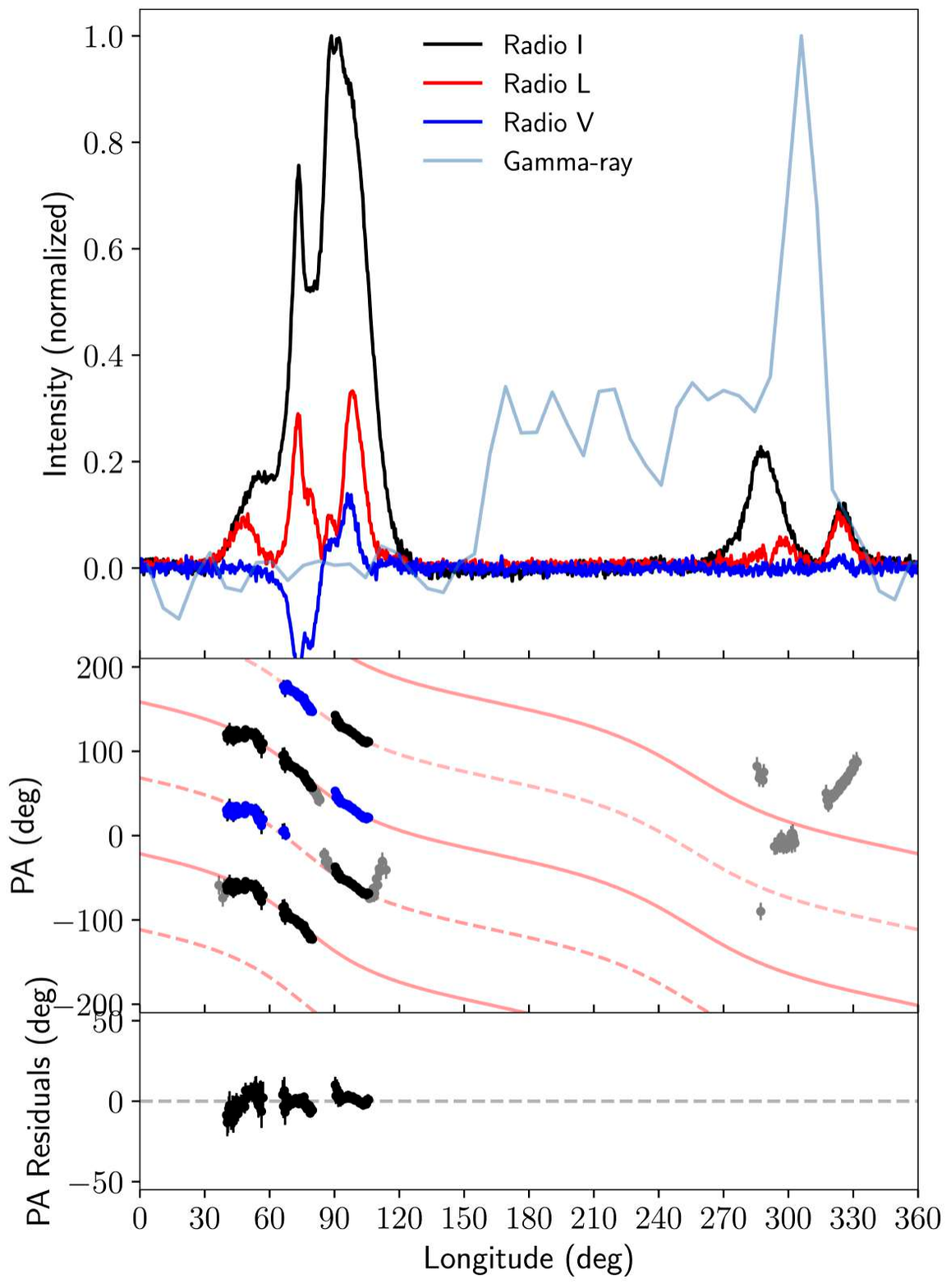}  \\
\end{center}
\caption{PSR~J1125$-$5825. The top panel shows the total intensity (black), linear polarization (red) and circular polarization (blue) for the radio emission. The light-blue line indicates the gamma-ray profile.  In the middle panel the PAs are shown along with the best fit RVM. PA values are shown as measured in black and offset by 90\degr\ in blue. The dashed lines correspond to a RVM solution separated in PA by 90\degr. PA values in gray have been not been modeled.  The bottom panel shows the residuals between model and data.}
\label{fig:1125}
\end{figure}

\begin{figure}
\begin{center}
\begin{tabular}{c}
\includegraphics[width=8cm]{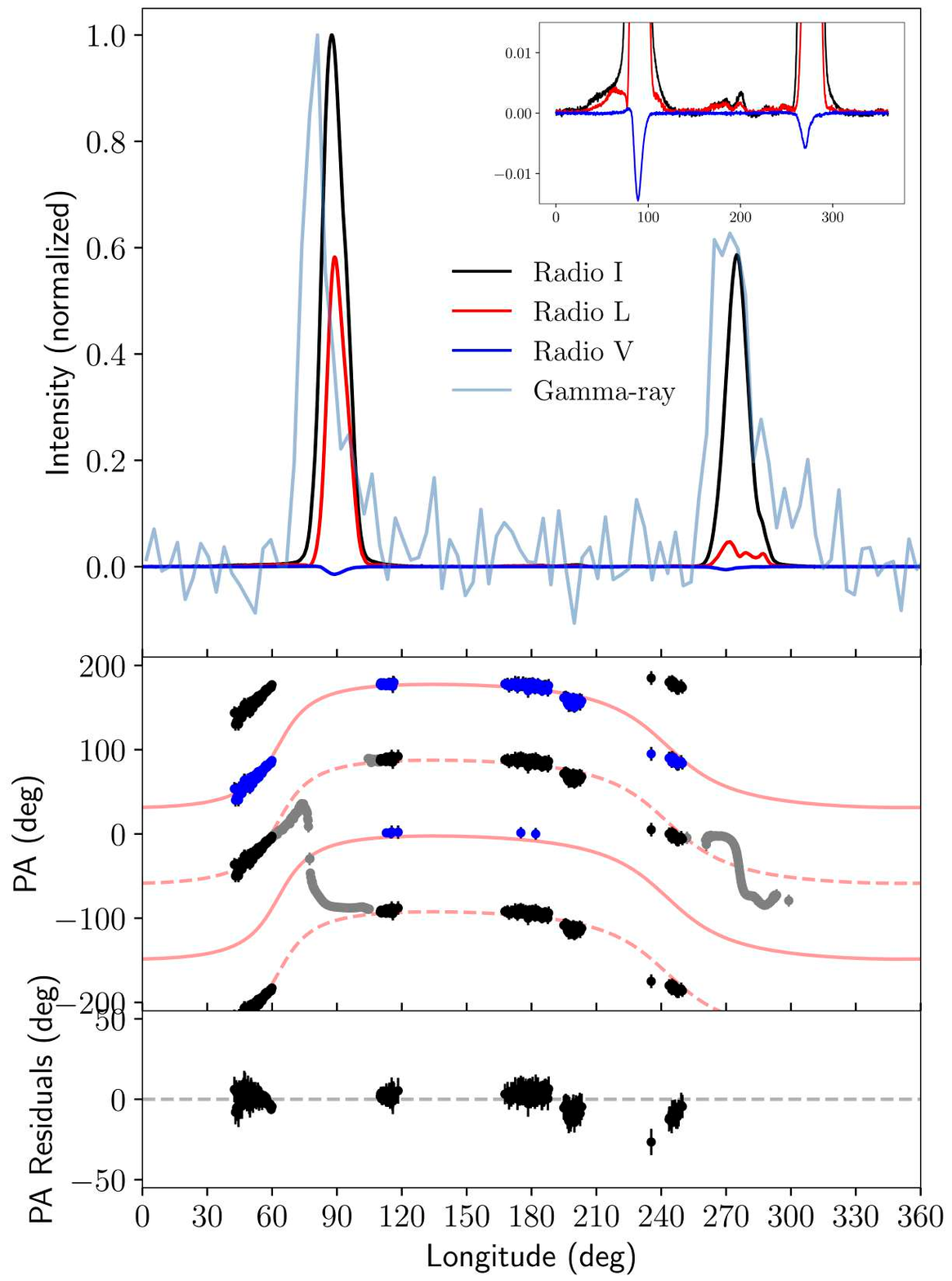}  \\
\end{tabular}
\end{center}
\caption{PSR~J1939+2134 is shown in a similar way to Figure~\ref{fig:1125}. The additional inset in the top panel top panel highlights the low-level emission.}
\label{fig:1939}
\end{figure}
\section{Discussion}
\subsection{The millisecond pulsars}
There are three key points from our inspection of MSP profiles. First, a large fraction (39\%) of the population have disjoint components in their profiles. Secondly, the class D pulsars have fast spin periods (median value 3.6~ms) and a high fraction are gamma-ray emitters. Thirdly, the disjoint radio components are, in many cases, aligned with gamma-ray components.

In the gamma-rays, much progress has been made recently in understanding the emission process (either synchrotron or curvature radiation) for producing gamma-rays in the current sheet beyond the light cylinder \citep{khk+19,kwh+22}. Most recently, \citet{pet24} and \citet{cfd25} have provided an atlas of gamma-ray profiles depending on the inclination angle of the rotation and magnetic axes ($\alpha$) and the viewing angle ($\zeta$).

It is our contention that there are two distinct regions of radio emission in MSPs. The first is from low in the magnetosphere above the polar cap (PC), as per the conventional picture. The second is co-located with regions of gamma-ray emission, beyond the light cylinder (LC) in the current sheet. We would therefore expect the LC components to be separated from the PC components by an amount which depends on the viewing geometry (see e.g. \citealt{pet24}) and that the LC radio components are in-phase with the gamma-ray emission. The LC components will have polarization properties distinct from the PC components; in particular the position angle of the linear polarization is unlikely to follow the RVM.

We illustrate our ideas with two examples. Figure~\ref{fig:1125} shows PSR~J1125$-$5825 which has a radio profile with two distinct regions. The weaker component aligns with the dominant gamma-ray component whereas the main radio component has no gamma-ray counterpart. We therefore assign the dominant radio component as coming from the polar cap, with the weaker component arising outside the light cylinder co-located with the gamma-rays. Fitting an RVM to the whole profile is indeed not possible, and hence, we applied the RVM to the strong component only. Ignoring a peculiar upturn of the PAs, possibly due to an unresolved orthogonal jump, we obtain $\alpha = 94\pm 13$\degr\ and $\zeta = 123\pm 12$\degr. Our second example is the original millisecond pulsar PSR~J1939+2134 (B1937+21) shown in Figure~\ref{fig:1939}. Long thought to be an orthogonal rotator, its PA swing is nevertheless very flat across both the bright radio components (when ignoring a resolved orthogonal jump). These radio components align perfectly with the gamma-ray profile. High sensitivity observations reveal additional, very low-level components that represent the PC emission in our picture. It is possible to fit an RVM to the PC components alone, resulting in $\alpha=108\pm2$\degr\ and $\zeta = 96\pm2$\degr.

\begin{figure}
\begin{center}
\includegraphics[width=9cm]{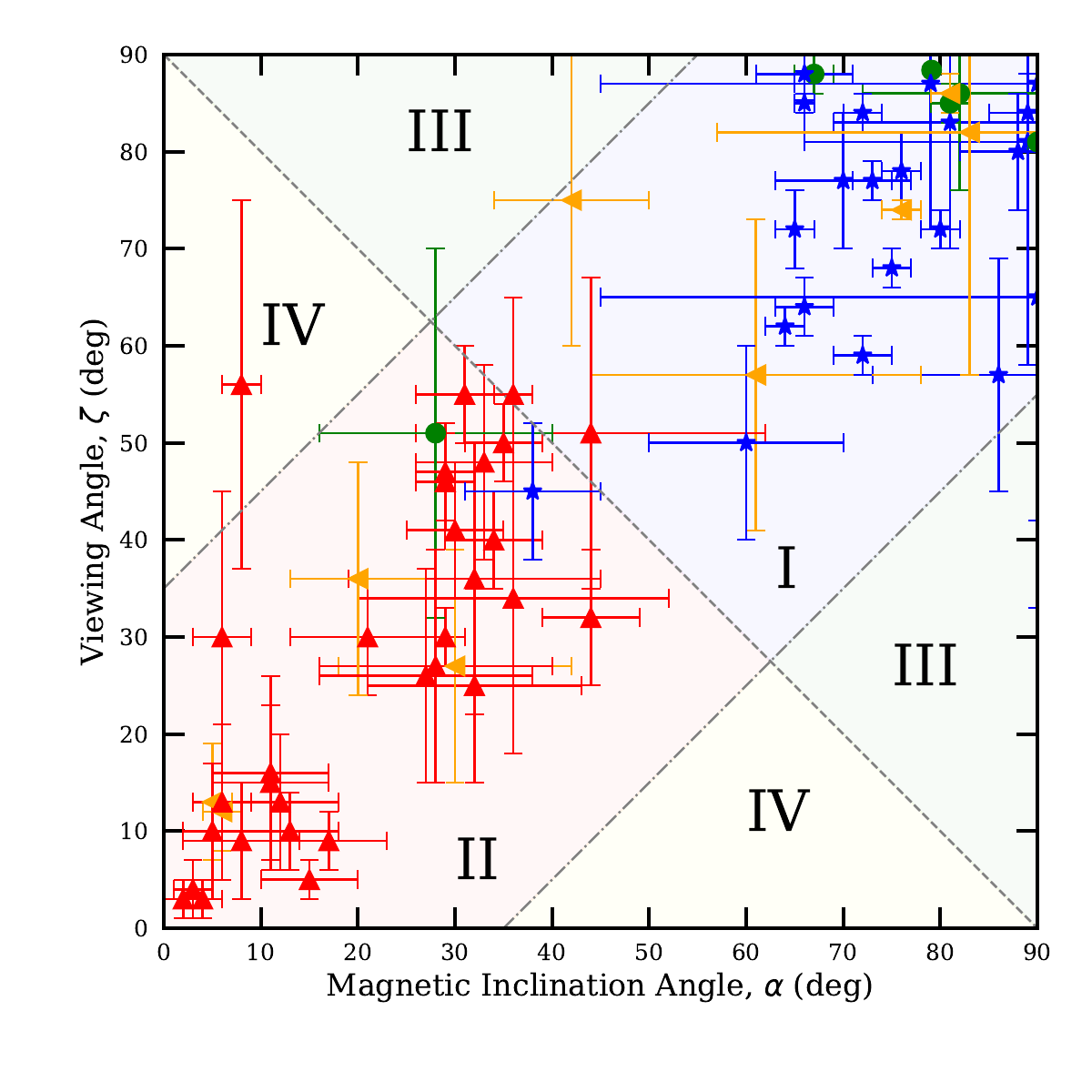}  
\end{center}
\caption{Regions of the $\alpha$-$\zeta$ plane showing detectability of emission regions. In the green regions only the LC components are visible, in the blue region both LC and PC components are visible. Only the PC emission is visible from the red region and in the yellow areas neither the LC nor PC emission is visible. A spin period of 3.5~ms and an emission height for the PC emission of 30~km has been assumed. The source geometries derived and given in the tables in the Appendix are shown: Sources from Table~\ref{tab:LC} as {\em blue}, from \ref{tab:NOLC} as {\em green}, from \ref{tab:DNO} as {\em orange}  and from \ref{tab:CLUM} as {\em red} symbols. Values of $\alpha$ that are larger than 90\degr\ have been folded into the plotted range.}
\label{fig:az}
\end{figure}

We further adopt the following idea based on the angles $\alpha$ and $\zeta$ as shown also in Figure~2 of \citet{pet24}. We take the median spin-period of the pulsars in our sample, 3.5~ms, and assume an emission height of 30~km. The half-opening angle of the emission beam is then 36.4\degr. This defines the dashed-dotted lines in Figure~\ref{fig:az}. \citet{pet24} showed the observer's visibility of the current sheet is defined as being above the dashed diagonal line in Figure~\ref{fig:az}. There are then three regions within the $\alpha$-$\zeta$ plane where emission can be detected as shown by the colour shading in Figure~\ref{fig:az}. The first (region I) is where both PC and LC emission can be seen, generally for $\alpha>45$\degr and $\alpha$ close to $\zeta$ (blue shading). The second (region II) is where only PC emission can be detected, with low values of $\alpha$ and/or low values of $\zeta$ (red shading). Finally, region III is where only LC emission can be seen, regions far from the $\alpha=\zeta$ line (green shading). The geometry of region IV is such that neither LC nor PC emission is detectable (yellow shading).

We can determine the relative fractions of potentially detectable pulsars in each of the shaded areas. As the (underlying) distribution of $\alpha$ is unknown, we assume (a) isotropic and (b) uniform. Results are shown in Table~\ref{tab:frac}. The table shows that, when considering both PC and LC regions, emission covers 85\% of the celestial sphere in the uniform case and 94\% in the isotropic case. The actual detectability of a pulsar depends on the luminosity of the PC and LC emission and the (often severe) selection effects of a given radio or gamma-ray observation.
\begin{table}
\caption{Fraction of pulsars in each of the areas shown in Figure~\ref{fig:az} for an isotropic (columns 2 and 3) and uniform (columns 4 and 5) distribution of $\alpha$.}
\label{tab:frac}
\begin{center}
\begin{tabular}{llll}
\hline   
\hline
Category & Region & Fraction & Fraction  \\
 & & Isotropic & Uniform \\
\hline
LC + PC & I & 0.60 & 0.44 \\
PC only & II & 0.15 & 0.22  \\
LC only & III & 0.19 & 0.19 \\
Total detectable & I+II+III & 0.94 & 0.85 \\
Total not detectable & IV & 0.06 &  0.15\\
\hline
\end{tabular}
\end{center}
\end{table}

We classify the pulsars in our sample according to their radio profiles (class D or class C), the presence (or not) of gamma-rays and the alignment (or not) between the radio and gamma-ray profiles. In addition we deduce the angles $\alpha$ and $\zeta$ from the RVM fitting where possible. Tables are given in Appendix A and a description of the pulsars in Appendix B.

Table~\ref{tab:ONLY} lists the 21 pulsars with radio LC emission and no PC emission (i.e. from the green regions of Figure~\ref{fig:az}). It is particularly striking that the RVM cannot be fitted for any of these pulsars. Table~\ref{tab:LC} lists the 40 pulsars with both radio PC and LC emission (blue region of Figure~\ref{fig:az}). In these cases we see radio components with both gamma-ray counterparts (the LC components) and without gamma-ray counterparts (the PC emission). Combining these two tables we see that at least 61 of the 81 gamma-ray MSPs show evidence for LC radio emission.

Table~\ref{tab:NOLC} shows the remaining 20 pulsars with gamma-ray emission which we have found difficult to classify. Some have very low gamma-ray counts and a reliable profile cannot be created. Some have overlapping radio and gamma components but it is unclear how to assign PC and LC emission. In summary therefore, we think it likely that nearly all gamma-ray MSPs show (or will show) components aligned with at least some part of the radio profile. This is a striking result.

Table~\ref{tab:DNO} shows the remaining pulsars with disjoint profiles. Given that many of the radio profiles look similar to those pulsars found in Table~\ref{tab:LC} we expect that both radio PC and LC emission are seen. However, none have (yet) been detected in gamma-rays. Some have not been folded to produce gamma-ray profiles due to poor radio ephemerides. For the rest we use the `heuristic' Fermi detection threshold of $G>10^{-12}$ computed via
\begin{equation}
    \rm{log}(G) = -11.1\,\,\,+\,\,\,0.5\rm{log}(\dot{E}/10^{33})\,\,\, -\,\,\,2log(d)
\end{equation}
(adapted from equation 25 of \citealt{3PC})
to determine whether or not the pulsars are potentially detectable. We see that 
12 of these pulsars are above the threshold. It is noticeable that the spin-down energies of the pulsars in this Table are smaller than those in Tables~\ref{tab:ONLY} to \ref{tab:NOLC}, perhaps indicating lower levels of gamma-ray luminosity.

Finally, Table~\ref{tab:CLUM} lists the pulsars with contiguous profiles that do not have gamma-ray emission, but are above the nominal Fermi threshold for detection. It is our premise that the majority of these pulsars have unfavourable viewing geometries for LC emission to be observed (i.e. red region of Figure~\ref{fig:az}).

Inspecting the $\alpha$-$\zeta$ distribution in Figure~\ref{fig:az} is intriguing. The vast majority of sources in Table~\ref{tab:LC}  (the blue symbols) fall within the region where both PC and LC emissions are expected to be detected (the blue shaded region I). Similarly, sources from Table~\ref{tab:CLUM} (red symbols) appear to have unfortunate viewing geometry, meaning they do not show gamma-ray emission despite being sufficiently energetic and nearby. We expect them in the red shaded region II and infer that these pulsars have mostly aligned geometries. The position of sources with unclear classification (Table~\ref{tab:NOLC}, green symbols) in Figure~\ref{fig:az} may imply that some of them have LC components.  Finally, the sources listed in Table~\ref{tab:DNO} (orange symbols) should have been detected in gamma rays, assuming that the detected disjoint components originate from close to the LC. The majority of these sources are indeed located in the blue region I. More sensitive radio observations of the remaining sources may reveal contiguous profiles without disjoint components. 
\begin{figure}
\begin{center}
\begin{tabular}{c}
\includegraphics[width=8cm]{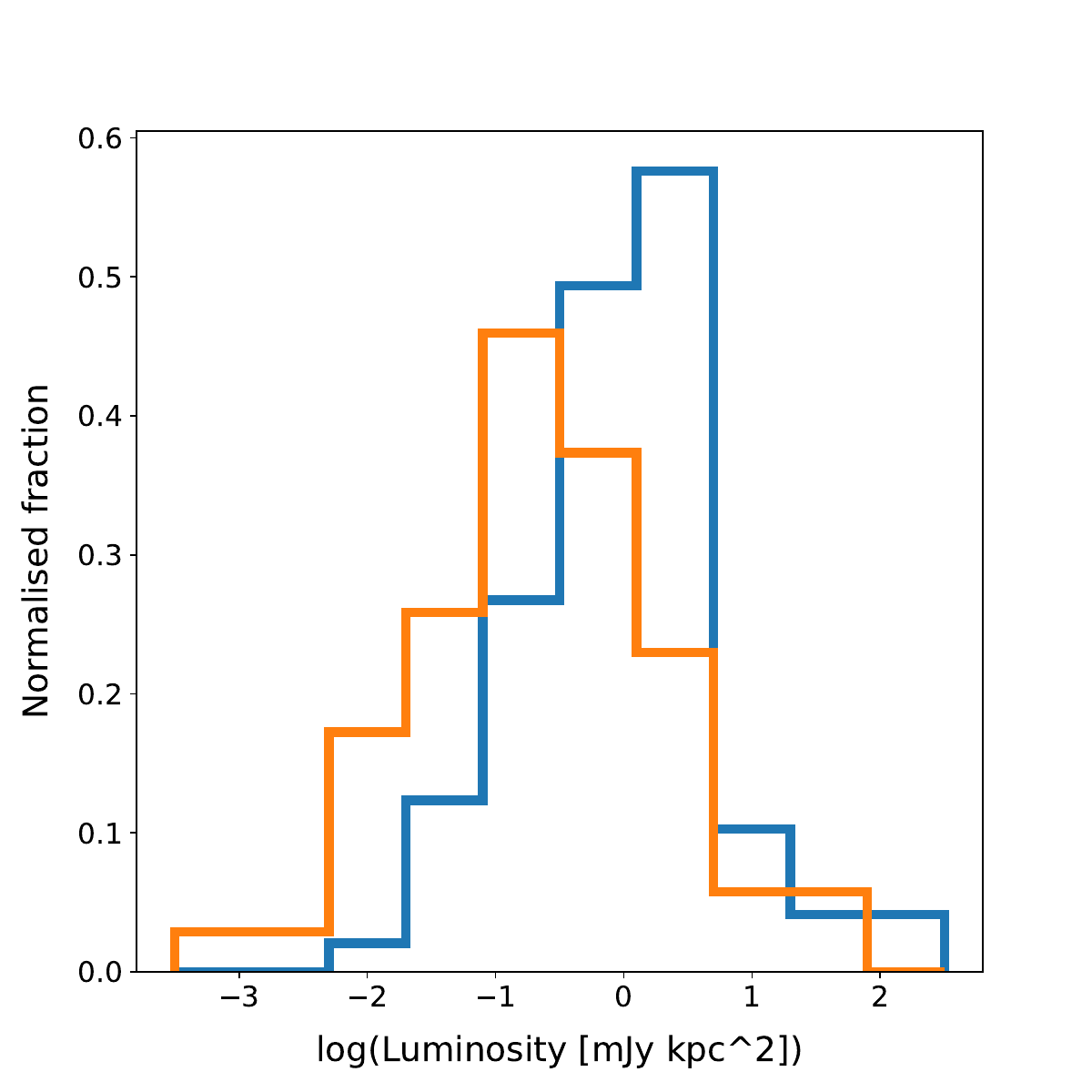}  \\
\end{tabular}
\end{center}
\caption{Luminosity histogram for the PC components (blue) and the LC components (orange) normalised to equal area.}
\label{fig:lumhist}
\end{figure}

We also emphasize the paucity of sources in the green (III) and yellow (IV) regions of Figure~\ref{fig:az}. We expect the sources from Table \ref{tab:ONLY} to be located in the green region, where only LC emission is expected. However, as we do not have reliable geometry information based on the radio observations, we cannot plot these sources. The fact that no other source falls into the green region and only one in the yellow region, where no emission is expected to be detected, strongly supports our findings. The one source that falls into the yellow region is PSR J0509$+$0856, albeit with large error bars. These are the results from the very unusual, very flat PA swing that shows some unexplained wiggles as discussed in Section~\ref{dis0509}.

We also compare the frequency with which we observe flat PA swings in components that we identify as possibly originating from the LC. For the pulsars listed in Tables A1 and A2, the percentages of flat and non-flat PAs, and of cases where this cannot be reliably determined, are as follows A1: (67\%/ 0\% / 33\%) and A2: (56\%/ 8\% / 36\%), whereas for A5:  (35\%/ 48\% / 27\%), respectively. In other words, although misclassifications and additional magnetospheric propagation effects cannot be ruled out entirely, flat PAs are much more likely to be found in LC components.

\begin{figure}
\begin{center}
\includegraphics[width=8cm]{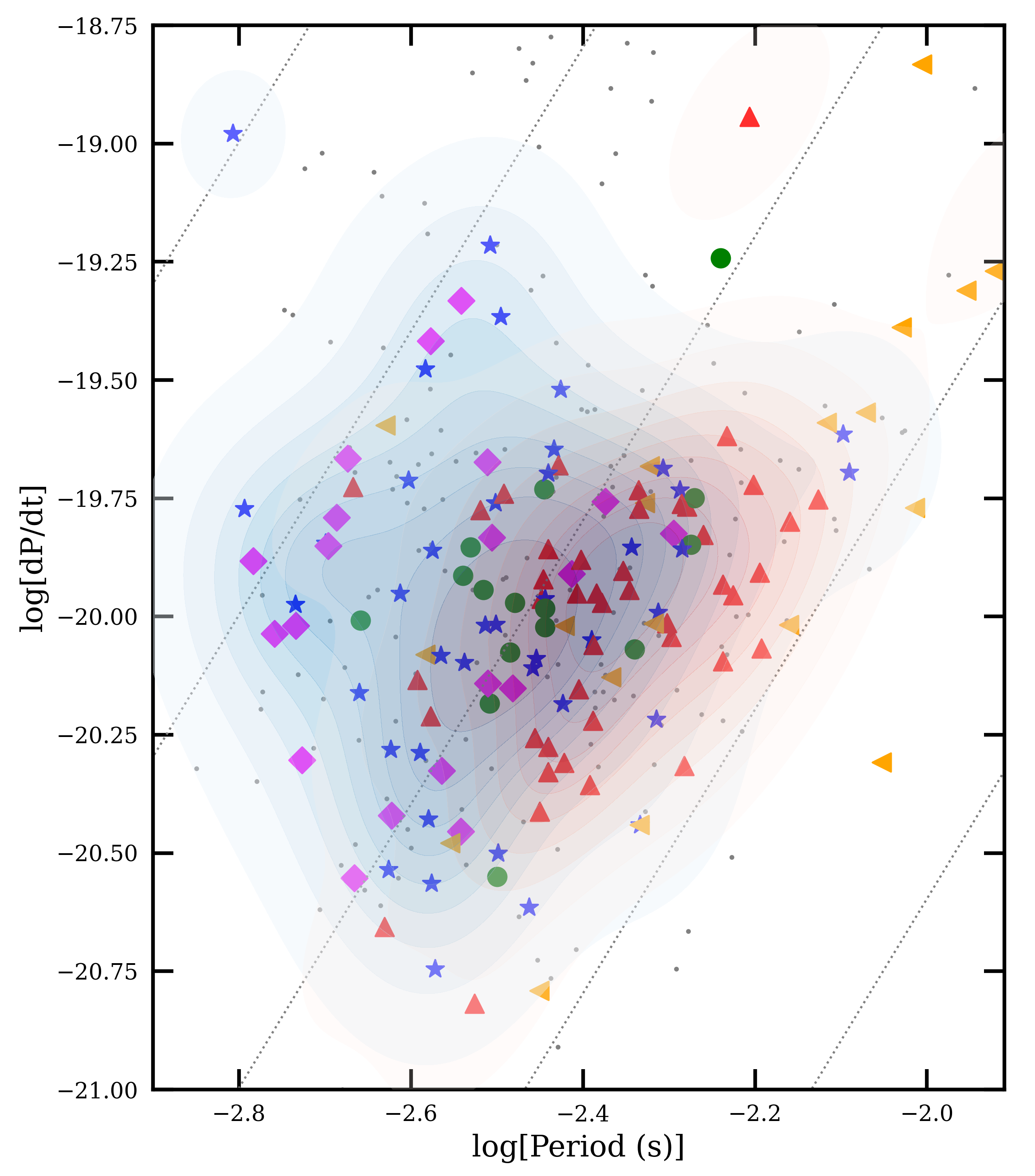}  
\end{center}
\caption{A $P$-$\dot{P}$-diagram showing the millisecond pulsar region. Dotted lines indicate constant $\dot{E}$.
Pulsars discussed here are shown with symbols in colours identical to Figure~\ref{fig:az}. Sources from Table~\ref{tab:LC} as {\em blue}, from \ref{tab:NOLC} as {\em green}, from \ref{tab:DNO} as {\em orange} and from \ref{tab:CLUM} as {\em red} symbols. Sources from Table~\ref{tab:ONLY} are shown in {\em magenta}. The shaded regions indicate densities of pulsars identified with LC (blue) and PC components (red).}
\label{fig:ppdot}
\end{figure}

We compare the luminosity of the LC components and the PC components in the following way. We define the luminosity via $Sd^2$ with $S$ the flux density at 1400 MHz and $d$ the distance obtained from the pulsar catalogue. We use the pulsars in Table~\ref{tab:ONLY} (LC only pulsars) and Table~\ref{tab:CLUM} (PC only pulsars). For the pulsars in Table~\ref{tab:LC} we compute the flux densities of the PC and LC components separately. Figure~\ref{fig:lumhist} shows the luminosity distributions. The luminosity of the LC components is a factor of 5 smaller than that of the PC components and has a broader distribution. Examination of the spin-down energies shows that the pulsars with only LC components have higher values of $\dot{E}$ than those with only PC components by a factor of 5. We therefore suggest that LC components have lower luminosities than their PC counterparts and that the luminosity is more strongly dependent on $\dot{E}$ making the LC components more difficult to detect for the lower $\dot{E}$ pulsars, although we caution that selection effects in radio pulsar surveys are also playing a role. The $P$-$\dot P$-diagram shown in Figure~\ref{fig:ppdot} illustrates some of these differences. Pulsars with fast spin periods and higher values of $\dot{E}$ are more likely to show LC components compared to the longer period, lower $\dot{E}$ counterparts although we caution that selection effects in both radio and gamma surveys also play a role in the location of pulsars in the Figure.

In the time since the \citet{3PC} catalogue, gamma-ray MSPs continue to be discovered, generally with low radio fluxes. The pattern remains similar to that described above. Of the 10 newly discovered MSPs with gamma-ray emission in \citet{bnc+24} and \citet{kjc+25}, at least 5 of them appear to have LC components judging from the figures contained in those papers. We note also that two isolated (non-binary) gamma-ray MSPs are known without radio counterparts down to a limiting flux density of $50\mu$Jy \citep{cdw+25}. We surmise that the geometry is such that the radio PC emission is not in our line of sight and that deeper radio searches may be warranted to detect faint LC components.

\subsection{Slow pulsars}
From our analysis of 1492 pulsars, there are 51 pulsars in Class D. Of these, 38 are interpulse pulsars already known in the literature \citep{jk19,serylak,swyw25}. We have discovered a further interpulse in PSR~J1740--3015. The interpulse emission is offset from the main pulse by 175\degr\ and its amplitude  is only 0.0035 that of the main pulse. Unfortunately, no polarization is detectable from the interpulse.

We can use a toy model to compute the number of interpulses expected in the population of slow pulsars. We assume the emission height to be 300~km, independent of period. This in turn yields a cone opening angle, $\rho$ (which does depend on the period). We also assume that $\alpha$ and $\zeta$ are isotropically distributed and do not evolve with time. We take the period distribution of the 1492 pulsars in our dataset and from this derive an interpulse fraction of 5.9\%. The interpulse fraction for the observed sample is only 2.6\%, substantially less than expected from the simple calculation above. This is in-line with more detailed calculations from the literature \citep{wj08,nbg+20} and likely implies that $\alpha$ evolves towards zero with time \citep{jk19b}, i.e. the assumption of an isotropic distribution of $\alpha$ in the evolved population is incorrect.

The 12 (non-interpulse) pulsars with disjoint emission are listed in Table~\ref{tab:slow} along with the ratio of the component amplitudes and the separation between the components. Figure~\ref{fig:slow} shows three examples (PSRs~J0401--7608, J1851+0418 and J1852--0118) of the detection of low-level components well separated from the main pulse emission. We know of only one other example in the literature: PSR~J0304+1932 which has a very low amplitude component located 120\degr\ from the main component \citep{swyw25}. We carefully examined the pre and post cursor pulsars described in \citet{bmr15} but none of them appear to be truly disjoint. 

In summary, excluding the obvious interpulse pulsars, it appears that less than 1\% of the slow pulsar population have disjoint components. PSR~J0631+0646 is the only pulsar in Table~\ref{tab:slow} that is a gamma-ray emitter, neither of its radio components align with the gamma-ray profile \citep{3PC}. The pulsars have no particular stand-out parameters; their spin periods, spin-down luminosities, characteristic ages and magnetic field strengths are very similar to many other slow pulsars. However, examination of Table~\ref{tab:slow} shows that the faster rotating pulsars have components separated by $<70$\degr, whereas the slower rotating pulsars have components separated by $>100$\degr. None of the pulsars have $P>650$~ms.

The origin of these discrete components is puzzling. Could they be from the LC, similar to the MSP components? If so, why these particular pulsars and not the known gamma-ray slow pulsars? Could they be almost aligned rotators with very patchy emission beams? If so, why are they mostly the faster rotating pulsars rather than the much slower pulsars which are more likely to be aligned rotators? Could they be evidence for some sort of multi-pole magnetic field structure? If so, why are these pulsars such a small fraction of the overall population? On current evidence these questions remain open.

\begin{table}
\caption{Slow pulsars with disjoint components (excluding interpulses) arranged in order of increasing spin period. The ratio of the amplitudes of the components are given in Column 3 and the separation in Column 4. The last column gives the reference for the profile: (1) \citet{pkj+23} (2) \citet{whx+23}.}
\label{tab:slow}
\begin{center}
\begin{tabular}{lllrrl}
\hline   
\hline
Name & Period & log($\dot{E}$) & Ratio & Delta & Ref\\
& (ms) & (ergs$^{-1}$) & & (deg) \\
\hline
J0631+0646 & 110.9 & 35.0 & 0.6 & 58 & (1) \\
J1047--6709 & 198.4 & 33.9 & 0.05 & 57 & (1) \\
J1904+0738 & 208.9 & 33.3 & 0.1 & 52 & (2) \\
J1905+0902 & 218.2 & 34.1 & 0.05 & 43 & (2) \\
J1603--2531 & 283.1 & 33.4 & $<0.01$ & 126 & (1)\\
J1851+0418 & 284.6 & 33.3 & 0.2 & 148 & (1) (2)\\
J1850+1335 & 345.5 & 33.1 & $<$0.01& 71 & (1) (2)\\
J1852--0118 & 451.5 & 32.9 & 0.6 & 136 & (1) (2)\\
J1017+3011 & 452.8 & 32.3 & 0.9 & 128 & (2) \\
J0401--7608 & 545.2 & 32.6 & $<0.01 $ & 92 & (1)\\
J1853+0259 & 585.5 & 31.3 & 0.4 & 105 & (2) \\
J1332--3032 & 650.4 & 31.9 & 0.4 & 107 & (1) \\
\hline
\end{tabular}
\end{center}
\end{table}

We also re-examined the 8 `classical' interpulses from our sample that have known gamma-ray emission. PSR~J1057--5226 (B1055--52) has a single broad gamma-ray component which is partly aligned with the `main pulse' component in the radio (as already reported by \citealt{abdo10}). Neither the shape of the gamma-ray profile or the alignment between the gamma and radio are expected for an orthogonal rotator. Could it be the case that the value of $\alpha$ is significantly less than 90\degr\ and that the main radio component is in fact light cylinder emission? This may alleviate some of the problems discussed by \citet{ww09} in mapping the beam of this pulsar. PSR~J1705--1906 (B1702-19) also has a single narrow gamma-ray component aligned with the `interpulse' component in the radio profile. We note the peculiarity that both these pulsars have been linked with phase-locked delays between the behaviours of the two components \citep{wws07,wwj12} something also seen in the light cylinder components of the MSP B1957+20 \citep{mkm+18}. Finally, although the s/n of the gamma-ray profile of PSR~J1935+2025 is low, it appears as if both gamma-ray component align with the two radio components leading to the tantalising possibility that only light cylinder emission is being seen in this slow pulsar. The recent detection of radio emission at a very low flux density in PSR~J0359+5414 \citep{dpe+25} shows the radio and gamma profiles are aligned. It could therefore be the case that radio LC emission is very weak, perhaps explaining the lack of LC components seen in the Vela pulsar at a level at least $10^4$ times below that of the main PC component.

\section{Summary}
The MSPs and the slow pulsar populations share many characteristics and behaviours in spite of their large differences in spin periods and magnetic field strengths. A striking difference between the integrated radio profiles of these two populations is the presence of disjoint components. We have shown that 35\% of MSP profiles have disjoint components as opposed to only 3\% of the slow pulsar profiles. 
Remarkably, these percentages have not changed since the initial findings by  \cite{paperI} , who determined 36\% for MSPs and 2\% for normal pulsars using a sample that was about ten times smaller (and more sensitivity limited).
Importantly, in this work, we find that the radio profiles of many of these pulsars have components that are aligned with the gamma-ray profiles. 
These components often exhibit a high degree of polarisation and typically demonstrate a flat position angle swing. We note that flat PAs are also often observed in fast radio bursts \cite[FRBs, e.g.][]{FRBsPA}, and one might speculate whether such emission could be related to emission beyond the light cylinder of a neutron star for these sources.

We therefore surmise that radio emission can be produced beyond the light cylinder in MSPs, in regions co-located with the production of gamma-rays. In addition, radio emission at low altitudes from above the polar caps is also produced. There is some evidence that the LC components have lower luminosities than their PC counterparts. We propose a simple geometrical explanation for the presence (or not) of light cylinder emission and the alignment (or not) between the radio components and the gamma-ray components. We also show that RVM fits to $\alpha$ and $\zeta$ are supportive of this idea. These findings help to explain some of the few peculiar differences in the emission properties of millisecond and normal pulsars that were noted early on \citep{paperI,paperII}. 

If this picture is correct, it has implications for theorists and observers alike. For the theorist the challenge is devising a way to produce coherent radio emission in the current sheet \citep{lyu19,pusc19}. Observationally, the beaming fraction of light cylinder emission is higher than the beaming fraction of polar cap emission meaning that more MSPs should be detectable than previously thought. This has consequences for modeling the Galactic population of millisecond pulsars and the relative sizes of the population of radio-loud and radio-quiet millisecond pulsars.

The polarization and single pulse properties of the light-cylinder components may be very different to those of conventional components. The presence of giant pulses in PSRs B1937+21 and B1957+20 in LC components may be an indicator of power-law statistics in these components more broadly. A study of PSR~B1957+20 by \citet{mkm+18} shows that mode-changing occurs in the components we have identified as LC emission in addition to giant pulses. The stability (or not) of the LC components may also have implications for MSP timing. When we examine the MSPs used in pulsar timing arrays (PTAs) we see that the pulsars with only LC components (Table~\ref{tab:ONLY}) are under-represented. Five of these pulsars are so-called `spider' pulsars \citep{spidercat} making them unsuitable for PTAs. Other reasons for their lack of representation could arise from the presence of higher jitter noise and/or higher timing rms in these pulsars and/pr a consequence of the lower luminosity of the LC components. We plan to examine these facets of the radio LC components in more detail in an upcoming work.

In the slow pulsar population there are hints that some of the pulsars previously classified as showing interpulses are instead made up of both polar cap and light cylinder radio emission, but the prevalence of such emission seems lower than that in the MSP population, perhaps due to low flux levels of the LC components.

\section*{Acknowledgements}
We thank D. Smith and C. Clark for answering our constant stream of questions regarding the gamma-ray population and the referee for helping improve the paper. We thank M. Burgay and A. Possenti for enabling us to present this work at the Sardinia Pulsar Conference and to our colleagues there for stimulating conversations.

\section*{Data Availability}
The data used in this paper are all publicly available from the references cited in the text.



\bibliographystyle{mnras}
\bibliography{msps} 

\appendix
\section{Table of Results}
In the tables below, the flags are as follows:
\begin{itemize}
    \item C Contiguous radio profiles
    \item D Disjoint radio profiles
    \item F Gamma-rays folded, not detected
    \item N Gamma-rays not folded, not detected
    \item L Above the nominal Fermi threshold for detection
\end{itemize}
The references for the tables are (1) \citet{sbm+22}, (2) \citet{wmg+22}, (3) \citet{whx+23}, (4) \citet{xjx+25}. We use \citet{3PC} to compare the phases of the gamma-ray and radio profiles. Pulsar periods (in ms) and spin-down energies ($\dot{E}$; ergs$^{-1}$) are taken from version 2.6.5 of the pulsar catalogue.

\begin{table}
\caption{Pulsars with only LC emission and no PC emission.}
\label{tab:ONLY}
\begin{center}
\begin{tabular}{llccc}
\hline
\hline
PSR & Ref & P & log($\dot{E}$) & Flag\\
    &     & (ms)   &                &     \\
\hline
J0034--0534 & 1.4. & 1.88 & 34.5 & C \\
J0340+4130 & 2.3.4 & 3.30 & 33.9 & C \\
J0605+3757 & 3.4. & 2.73 & 34.0 & D \\
J0610--2100 & 1. & 3.86 & 33.9 & D \\
J0636+5128 & 2.4. & 2.87 & 33.8 & C \\
J1035--6720 & 1. & 2.87 & 34.9 & D \\
J1142+0119 & 1.3. & 5.07 & 33.7 & D \\
J1312+0051 & 1.3. & 4.23 & 34.0 & D \\
J1400--1431 & 1. & 3.08 & 34.0 & C \\
J1431--4715 & 1. & 2.01 & 34.8 & C \\
J1513--2550 & 1. & 2.12 & 35.0 & D \\
J1536--4948 & 1. & 3.08 & 34.5 & D \\
J1543--5149 & 1. & 2.06 & 34.9 & C \\
J1747--4036 & 1.2. & 1.65 & 35.1 & C \\
J1816+4510 & 3. & 3.19 & 34.7 & D \\
J1843--1113 & 1.4. & 1.85 & 34.8 & D \\
J1858--2216 & 1. & 2.38 & 34.0 & C \\
J1902--5105 & 1. & 1.74 & 34.8 & D \\
J1921+1929 & 1.3. & 2.65 & 34.9 & D \\
J2051+50 & 3. & 1.68 & 0.0 & C \\
J2214+3000 & 4. & 3.12 & 34.3 & D \\
\hline
\end{tabular}
\end{center}
\end{table}
\begin{table}
\caption{Pulsars with both LC emission and PC emission.}
\label{tab:LC}
\begin{center}
\begin{tabular}{llccccc}
\hline
\hline
PSR & Ref & P & log($\dot{E}$) & Flag & $\alpha$ & $\zeta$ \\
    &     & (ms)   &                &      & (deg)    & (deg)   \\
\hline
J0030+0451 & 1.4. & 4.87 & 33.5 & D & 66 $\pm$ 3 & 64 $\pm$ 3 \\
J0101--6422 & 1. & 2.57 & 34.1 & D & & \\
J0154+1833 & 1.4. & 2.36 & 33.9 & D & 70 $\pm$ 7 & 77 $\pm$ 7 \\
J0613--0200 & 1.4.2 & 3.06 & 34.1 & C & & \\
J0614--3329 & 1. & 3.15 & 34.3 & D & 116 $\pm$ 2 & 118 $\pm$ 2 \\
J0751+1807 & 1.4. & 3.48 & 33.9 & D & 114 $\pm$ 5 & 92 $\pm$ 6 \\
J0931--1902 & 1.2. & 4.64 & 33.1 & D & & \\
J0955--6150 & 1. & 2.00 & 34.8 & D & 73 $\pm$ 2 & 77 $\pm$ 2 \\
J1024--0719 & 1.4.2 & 5.16 & 33.7 & D & & \\
J1036--8317 & 1. & 3.41 & 0.0 & D & 99 $\pm$ 12 & 97 $\pm$ 13 \\
J1125--5825 & 1. & 3.10 & 34.9 & D & 94 $\pm$ 13 & 123 $\pm$ 12 \\
J1125--6014 & 1. & 2.63 & 33.9 & D & 89 $\pm$ 4 & 84 $\pm$ 4 \\
J1207--5050 & 1. & 4.84 & 33.3 & D & & \\
J1231--1411 & 1. & 3.68 & 34.3 & D & 89 $\pm$ 23 & 81 $\pm$ 23 \\
J1301+0833 & 3. & 1.84 & 34.8 & C & & \\
J1302--3258 & 1. & 3.77 & 33.7 & D & 101 $\pm$ 14 & 93 $\pm$ 15 \\
J1327--0755 & 1. & 2.68 & 33.6 & D & & \\
J1455--3330 & 1.2. & 7.99 & 33.3 & D & 105 $\pm$ 2 & 112 $\pm$ 2 \\
J1614--2230 & 1.2. & 3.15 & 34.1 & D & 90 $\pm$ 45 & 87 $\pm$ 45 \\
J1658--5324 & 1. & 2.44 & 34.5 & D & & \\
J1730--2304 & 1. & 8.12 & 33.2 & D & 60 $\pm$ 10 & 50 $\pm$ 10 \\
J1741+1351 & 1.3.4 & 3.75 & 34.4 & D & 108 $\pm$ 3 & 121 $\pm$ 2 \\
J1744--1134 & 1.4.2 & 4.07 & 33.7 & D & 114 $\pm$ 1 & 95 $\pm$ 1 \\
J1745+1017 & 1.3.4 & 2.65 & 33.8 & D & 88 $\pm$ 6 & 80 $\pm$ 6 \\
J1811--2405 & 1. & 2.66 & 34.5 & C & & \\
J1832--0836 & 1.4.2 & 2.72 & 34.2 & D & 100 $\pm$ 2 & 72 $\pm$ 2 \\
J1855--1436 & 1. & 3.59 & 34.0 & D & & \\
J1921+0137 & 1.3. & 2.50 & 34.7 & D & & \\
J1939+2134 & 1.2.3 & 1.56 & 36.0 & D & 108 $\pm$ 2 & 96 $\pm$ 2 \\
J1946+3417 & 3.4. & 3.17 & 33.6 & D & 90 $\pm$ 45 & 65 $\pm$ 32 \\
J1959+2048 & 1.3. & 1.61 & 35.2 & D & 104 $\pm$ 2 & 78 $\pm$ 4 \\
J2017+0603 & 1.3.4 & 2.90 & 34.1 & D & & \\
J2042+0246 & 1.3. & 4.53 & 33.8 & D & 142 $\pm$ 7 & 135 $\pm$ 7 \\
J2043+1711 & 1.4. & 2.38 & 34.2 & C & 115 $\pm$ 2 & 108 $\pm$ 4 \\
J2124--3358 & 1. & 4.93 & 33.8 & D & & \\
J2150--0326 & 1.4. & 3.51 & 33.9 & D & & \\
J2215+5135 & 3. & 2.61 & 34.9 & D & & \\
J2234+0944 & 1.3.4 & 3.63 & 34.2 & D & & \\
J2241--5236 & 1. & 2.19 & 34.4 & D & & \\
J2302+4442 & 2.4. & 5.19 & 33.6 & C & & \\
\hline
\end{tabular}
\end{center}
\end{table}
\begin{table}
\caption{Pulsars for which the LC/PC classification is unclear}
\label{tab:NOLC}
\begin{center}
\begin{tabular}{llccccc}
\hline
\hline
PSR & Ref & P & log($\dot{E}$) & Flag & $\alpha$ & $\zeta$ \\
    &     & (ms)   &                &      & (deg)    & (deg)   \\
\hline
J0023+0923 & 1.3.4 & 3.05 & 34.2 & C & & \\
J0437--4715 & 1. & 5.76 & 34.1 & C & & \\
J0653+4706 & 3. & 4.76 & 33.9 & D & & \\
J0737--3039A & 1. & 22.70 & 33.8 & D & 79.1 $\pm$ 0.1 & 88.4 $\pm$ 0.1 \\
J0740+6620 & 2. & 2.89 & 34.3 & D & 67 $\pm$ 2 & 88 $\pm$ 2 \\
J1012--4235 & 1. & 3.10 & 33.9 & C & & \\
J1446--4701 & 1. & 2.19 & 34.6 & C & & \\
J1514--4946 & 1. & 3.59 & 34.2 & C & 82 $\pm$ 10 & 86 $\pm$ 10 \\
J1544+4937 & 3. & 2.16 & 34.0 & C & & \\
J1600--3053 & 1.2. & 3.60 & 33.9 & C & 152 $\pm$ 12 & 129 $\pm$ 19 \\
J1630+3734 & 3.4. & 3.32 & 34.1 & D & 90 $\pm$ 1 & 81 $\pm$ 1 \\
J1640+2224 & 1.4. & 3.16 & 33.5 & C & & \\
J1713+0747 & 1.2.4 & 4.57 & 33.5 & C & & \\
J1732--5049 & 1. & 5.31 & 33.6 & C & & \\
J1857+0943 & 1.3.4 & 5.36 & 33.7 & D & & \\
J1903--7051 & 1. & 3.60 & 33.9 & C & 105 $\pm$ 2 & 86 $\pm$ 2 \\
J1909--3744 & 1.2. & 2.95 & 34.3 & D & 81 $\pm$ 2 & 85 $\pm$ 2 \\
J1946--5403 & 1. & 2.71 & 0.0 & C & & \\
J2039--3616 & 1. & 3.28 & 34.0 & C & & \\
J2317+1439 & 1.4. & 3.45 & 33.4 & C & & \\
\hline
\end{tabular}
\end{center}
\end{table}
\begin{table}
\caption{Pulsars with disjoint profiles for which gamma--ray emission is expected but not yet detected}
\label{tab:DNO}
\begin{center}
\begin{tabular}{llccccc}
\hline
\hline
PSR & Ref & P & log($\dot{E}$) & Flag & $\alpha$ & $\zeta$ \\
    &     & (ms)   &                &      & (deg)    & (deg)   \\
\hline
J0406+3039 & 4. & 2.61 & 34.3 & F D L & & \\
J0645+5158 & 2.4. & 8.85 & 32.4 & F D & & \\
J0824+0028 & 1.4. & 9.86 & 33.8 & F D L & 174 $\pm$ 2 & 168 $\pm$ 4 \\
J0900--3144 & 1. & 11.11 & 33.1 & F D L & 5 $\pm$ 2 & 13 $\pm$ 6 \\
J0921--5202 & 1. & 9.68 & 32.9 & N D & & \\
J1216--6410 & 1. & 3.54 & 33.1 & N D L & 61 $\pm$ 17 & 57 $\pm$ 16 \\
J1529--3828 & 1. & 8.49 & 33.2 & N D & & \\
J1547--5709 & 1. & 4.29 & 33.6 & F D L & 42 $\pm$ 8 & 75 $\pm$ 15 \\
J1603--7202 & 1. & 14.84 & 32.3 & F D & 30 $\pm$ 12 & 27 $\pm$ 12 \\
J1618--3921 & 1. & 11.99 & 33.1 & F D & & \\
J1709+2313 & 1.3. & 4.63 & 33.1 & F D L & & \\
J1731--1847 & 1. & 2.34 & 34.9 & F D L & & \\
J1804--2717 & 1. & 9.34 & 33.3 & F D L & 160 $\pm$ 7 & 144 $\pm$ 12 \\
J1826--2415 & 1. & 4.70 & 33.8 & F D L & & \\
J1906+0055 & 1. & 2.79 & 33.8 & N D & & \\
J1918--0642 & 1.4.2 & 7.65 & 33.4 & F D L & & \\
J1923+2515 & 1.4. & 3.79 & 33.8 & F D L & 99 $\pm$ 2 & 94 $\pm$ 2 \\
J1935+1726 & 3. & 4.20 & 0.0 & N D & & \\
J2053+4650 & 3. & 12.59 & 33.5 & F D L & 104 $\pm$ 2 & 106 $\pm$ 1 \\
J2145--0750 & 1.4.2 & 16.05 & 32.4 & F D & & \\
J2236--5527 & 1. & 6.91 & 33.0 & N D L & 97 $\pm$ 26 & 82 $\pm$ 25 \\
J2322+2057 & 1.4. & 4.81 & 33.5 & F D L & & \\
\hline
\end{tabular}
\end{center}
\end{table}
\begin{table}
\caption{Class C pulsars above the heuristic Fermi detection level but no gamma--rays detected likely due to unfavourable viewing geometry}
\label{tab:CLUM}
\begin{center}
\begin{tabular}{llccccc}
\hline
\hline
PSR & Ref & P & log($\dot{E}$) & Flag & $\alpha$ & $\zeta$ \\
    &     & (ms)   &                &      & (deg)    & (deg)   \\
\hline
J0509+0856 & 1.4. & 4.06 & 33.4 & C & 8 $\pm$ 2 & 56 $\pm$ 19 \\
J0557+1550 & 1. & 2.56 & 34.2 & C & & \\
J0711--6830 & 1. & 5.49 & 33.6 & C & 6 $\pm$ 3 & 30 $\pm$ 15 \\
J0732+2314 & 1.4. & 4.09 & 33.5 & C & 29 $\pm$ 2 & 30 $\pm$ 3 \\
J1012+5307 & 2.4. & 5.26 & 33.7 & C & & \\
J1017--7156 & 1. & 2.34 & 33.8 & C & & \\
J1045--4509 & 1. & 7.47 & 33.2 & C & 147 $\pm$ 7 & 132 $\pm$ 10 \\
J1300+1240 & 1. & 6.22 & 34.3 & C & 32 $\pm$ 13 & 36 $\pm$ 14 \\
J1421--4409 & 1. & 6.39 & 33.3 & C & & \\
J1453+1902 & 1.3.4 & 5.79 & 33.4 & C & 35 $\pm$ 4 & 50 $\pm$ 4 \\
J1537--5312 & 1. & 6.93 & 33.3 & C & 44 $\pm$ 18 & 51 $\pm$ 16 \\
J1552--4937 & 1. & 6.28 & 33.5 & C & 32 $\pm$ 11 & 25 $\pm$ 10 \\
J1643--1224 & 1.4.2 & 4.62 & 33.9 & C & 167 $\pm$ 5 & 170 $\pm$ 4 \\
J1653--2054 & 1. & 4.13 & 33.8 & C & 169 $\pm$ 6 & 164 $\pm$ 10 \\
J1708--3506 & 1. & 4.50 & 33.7 & C & 31 $\pm$ 5 & 55 $\pm$ 5 \\
J1710+4923 & 3.4. & 3.22 & 34.3 & C & 4 $\pm$ 2 & 3 $\pm$ 2 \\
J1719--1438 & 1. & 5.79 & 33.2 & C & 168 $\pm$ 6 & 167 $\pm$ 7 \\
J1721--2457 & 1. & 3.50 & 33.7 & C & 44 $\pm$ 5 & 32 $\pm$ 7 \\
J1738+0333 & 1.3.4 & 5.85 & 33.7 & C & 159 $\pm$ 8 & 150 $\pm$ 6 \\
J1751--2857 & 1. & 3.92 & 33.9 & C & 28 $\pm$ 12 & 27 $\pm$ 12 \\
J1801--1417 & 1. & 3.62 & 33.6 & C & 169 $\pm$ 6 & 165 $\pm$ 8 \\
J1802--2124 & 1. & 12.65 & 33.1 & C & 144 $\pm$ 16 & 146 $\pm$ 16 \\
J1813--2621 & 1. & 4.43 & 33.8 & C & & \\
J1828+0625 & 1.3. & 3.63 & 33.6 & C & 15 $\pm$ 5 & 5 $\pm$ 2 \\
J1844+0115 & 1.3. & 4.19 & 33.8 & C & & \\
J1850+0124 & 1. & 3.56 & 34.0 & C & & \\
J1853+1303 & 1.3.4 & 4.09 & 33.7 & C & 29 $\pm$ 3 & 46 $\pm$ 5 \\
J1903+0327 & 1.3.4 & 2.15 & 34.9 & C & & \\
J1905+0400 & 3. & 3.78 & 33.6 & C & & \\
J1910+1256 & 1.3.4 & 4.98 & 33.5 & C & 172 $\pm$ 6 & 171 $\pm$ 6 \\
J1911--1114 & 1.4. & 3.63 & 34.1 & C & 144 $\pm$ 2 & 141 $\pm$ 2 \\
J1911+1347 & 3.4. & 4.63 & 33.8 & C & 36 $\pm$ 2 & 55 $\pm$ 10 \\
J1928+1245 & 3. & 3.02 & 34.4 & C & & \\
J1929+0132 & 3. & 6.42 & 33.1 & C & & \\
J1933--6211 & 1. & 3.54 & 33.5 & C & 30 $\pm$ 5 & 41 $\pm$ 7 \\
J1937+1658 & 1.3. & 3.96 & 33.9 & C & & \\
J1944+0907 & 1.3.4 & 5.19 & 33.7 & C & 17 $\pm$ 6 & 9 $\pm$ 3 \\
J2010--1323 & 1.4.2 & 5.22 & 33.1 & C & 7 $\pm$ 4 & 6 $\pm$ 4 \\
J2019+2425 & 1.3.4 & 3.94 & 33.7 & C & 3 $\pm$ 2 & 4 $\pm$ 3 \\
J2022+2534 & 4. & 2.65 & 34.1 & C & 2 $\pm$ 2 & 3 $\pm$ 2 \\
J2033+1734 & 1.3.4 & 5.95 & 33.3 & C & 146 $\pm$ 5 & 140 $\pm$ 5 \\
J2129--5721 & 1. & 3.73 & 34.2 & C & 29 $\pm$ 3 & 47 $\pm$ 5 \\
J2144--5237 & 1. & 5.04 & 33.4 & C & 27 $\pm$ 11 & 26 $\pm$ 11 \\
J2229+2643 & 1.3.4 & 2.98 & 33.4 & C & 5 $\pm$ 3 & 10 $\pm$ 7 \\
J2234+0611 & 1.3.4 & 3.58 & 34.0 & C & 6 $\pm$ 3 & 13 $\pm$ 8 \\
\hline
\end{tabular}
\end{center}
\end{table}

\section{Description of pulsar profiles and RVM fitting}
\subsection{Description of pulsars profiles from Table~\ref{tab:ONLY}}
These pulsars have radio LC emission aligned with the gamma-ray emission and no PC radio emission is detected.

\subsubsection{J0034-0534}
The radio profile is perfectly aligned with the gamma-ray profile. Only the weak, narrow post-cursor component is not aligned. This component has the highest degree of polarisation. The PA swing has three flat parts and only in the first part is some downward slope visible. It is not RVM like.

\subsubsection{J0340+4130}
The profile is nearly 190\degr\ wide and no disjoint components are detected. The second component of the double-component gamma-ray profile aligns with the radio component. The PA is flat and does not provide constraints.

\subsubsection{J0605+3757}
The radio profile consists of a strong double-peaked feature and a weak disjoint, trailing component with little polarization. The wide gamma-ray profile overlaps largely with both the strong radio component and the weak component. 
The PA is relatively flat at first, rising later on, making an RVM fit difficult.
If all the radio components are from the LC as suggested by the alignment the RVM is likely not the right description.

\subsubsection{J0610$-$2100}
Two components of almost equal width ($\sim$80\degr) are visible, the significantly weaker component trailing the stronger one by 140\degr\ (as measured from their centre). The gamma-ray profile covers most of the pulse period. It consists of a stronger and weaker component that appear to be connected, such that both radio components align with gamma-ray emission. This suggests that we see only LC components, supported by flat PA behaviour at the leading part of the strong radio component and associated with the complete weaker components. An RVM fit is therefore difficult.

\subsubsection{J0636+5128}
The radio profile is relatively narrow and consists of a single peak. The gamma-ray profile has low S/N but overlaps with the radio component. An RVM cannot be fitted.

\subsubsection{J1035$-$6720}
The S/N of the profile is low but shows a prominent gaussian-like feature and a weak component trailing by about 210\degr. The two gamma-ray peaks are somewhat misaligned with both radio components  but the high-energy profile is wide and therefore appears to overlap almost completely with the radio profiles. The limited S/N and the flat PA makes a RVM fit impossible. It is possible that with better S/N both components are connected, which would result in a profile width of about 210\degr.

\subsubsection{J1142+0119}
The profile shows a dominant component following 150\degr\ later by a weaker component consisting of three sub-components. The PA is defined over a wide range of longitudes, but it is very complex. 
FAST sensitivity suggests that all components are connected, forming a 260\degr-wide profile. 
The gamma-ray profile aligns with the strong radio component. The PA is mostly very flat, only the trailing part of the strong component shows some downward trend, but it is not sufficient to do a fit.

\subsubsection{J1312+0051}
The profile shows three features, although the dominant two may be connected by bridge emission. The weaker, narrow ($\sim$20\degr) component precedes the two dominant ones by 80\degr. Only the dominant radio feature is linearly polarised, resolving into three clear components. These are connected by a straight continuous, relatively flat PA. Some PA values are defined in the earlier components but the overall PA swing does not resemble an RVM. The gamma-ray profile has two separate wide peaks. One aligns with the two adjacent radio features, whereas the stronger (180\degr) wide  gamma-ray component's peak aligns with the weaker disjoint radio component.

\subsubsection{J1400$-$1431}
The radio profile is 66\degr\ wide, while the gamma-ray profile is about 180\degr\ wide and aligned with the radio profile. The profile has no visible polarisation.

\subsubsection{J1431$-$4715}
The radio profile extends over 160\degr\ of longitude; the trailing component is the brightest and has the highest degree of polarization. The gamma-ray profile has low S/N but appears to overlap with the radio profile. The PA swing is mostly flat after taking a prominent orthogonal jump into account. No RVM fit is possible.

\subsubsection{J1513$-$2550}
The profile shows two equally strong, equally wide ($\sim$50\degr) components, separated by about 130\degr. The gamma-ray profile looks nearly identical and aligns perfectly with the radio profile. The radio profile is only weakly polarised and therefore no PA information is available. Higher S/N may eventually show that both components are connected, which would result in a profile width of about 200\degr.

\subsubsection{J1536$-$4948}
The profile shows a two prominent components of nearly equal height, separated by about 75\degr. With the given S/N, the components are about 40\degr\ and 55\degr\ wide, respectively. There is a third disjoint, much weaker component, trailing by about 120\degr\ and with a width of about 50\degr. Given the low S/N it is difficult to discern if the components are connected or not. In such a case, the profile would be 250\degr\ wide. None of the components show significant polarisation. The gamma-ray profile is complex with four peaks. Its large width overlaps with most of the radio profile.

\subsubsection{J1543-5149}
The radio profile is just above 200\degr\ wide with very little polarisation. The PA that is detectabe is flat and non-RVM like. The simple gamma-ray profile is almost as wide and aligned with the radio profile.

\subsubsection{J1747$-$4036}
The profile is strange-looking and covers about 280\degr\ of longitude. A relatively narrow Gaussian-like component is followed by an almost equally strong radio component that is about 180\degr\ wide. The PA is very flat over 250\degr\ of longitude and not RVM like. The single-peaked gamma-ray profile is relatively narrow and sits between the two radio components. However, its S/N is low and gamma-ray emission may well extend over the full period, as suggested by the very flat radio PA.

\subsubsection{J1816+4510}
The profile shows one main gaussian-like component and a wide double-peaked component. All components have similar width, the main one has 65\degr, the first low-level component about 100\degr\ later has 55\degr\ width, and the third component, yet another 90\degr\ later, about 60\degr. The double-wing gamma-ray profile is perfectly aligned with the third weak component and the strong one. Only the first low-level component is clearly offset from the gamma-ray emission. A reliable RVM fit is not possible.

\subsubsection{J1843$-$1113}
The profile suggests the presence of two disjoint components, in addition to the prominent and relatively simple strong radio feature. It is possible that the two weaker components are connected and, with better S/N, even perhaps all components. The peak of the gamma-ray profile aligns with the strong radio peak. Overall, the gamma-ray emission is weak but may extend over the same 210\degr\ range as the radio profile. This would be consistent with the very flat PA that is only interrupted by an orthogonal jump between the first two components and the last one. An RVM fit is not possible.

\subsubsection{J1858$-$2216}
The radio profile consists of a single Gaussian with a width of about 80\degr. The PA swing is smooth and but relatively shallow and not sufficiently define to perform a meaningful RVM fit. The radio component aligns with the trailing part of the double gamma-ray profile.

\subsubsection{J1902$-$5105 }
This source shows two prominent radio feature, each about 75\degr\ wide. While the centres of the components are 210\degr\ apart, the strongest radio peak is 180\degr\ separated from the first component of the weaker feature. The gamma-ray profile is wide and shows two prominent peaks that align perfectly with the radio features. There is gamma-ray emission also in between these components, and indeed, there is a hint of additional radio emission there also. This would make the profile about 280\degr\ wide. The degree of polarisation is low and the flat PA is only defined in the strong component. An RVM fit is not possible.

\subsubsection{J1921+1929}
A strong radio feature is preceded by a weak disjoint component by about 80\degr. Neither of them is highly polarised but the PA is defined over a wide range. The gamma-ray profile has low S/N but appears to have two components. One appears somewhat offset to the weaker radio component. In contrast, the other gamma-ray feature largely overlaps with the radio strong component. An RVM fit is not possible.

\subsubsection{J2051+50}
The radio profile is relatively narrow and consists of two peaks, but the true extent may be larger with better S/N. The linear polarization is very high and the PA swing almost completely flat. The gamma-ray profile overlaps exactly the radio profile. 

\subsubsection{J2214+3000}
The radio profile shows a simple strong feature and a weaker, multi-component feature about 180\degr\ apart. There is emission trailing the strong component, connecting it to the weak one. No such emission is seen on the leading side. The separation of the components may, at first glance, suggests a main pulse-interpulse configuration. But the additional components and the observation that the gamma-ray emission apparently covers the whole pulse period, with two gamma-ray peaks aligned with the radio components, suggest that we only see LC emission. There is a low-level component which trails the main peak by about 100\degr. Overall, the PA swing is flat and non-RVM like. A RVM cannot be fitted. The components are at least 100\degr\ wide.

\subsection{Descriptions of pulsar profiles from Table~\ref{tab:LC}}
These pulsars have both radio LC and PC emission along with detected gamma-rays.

\subsubsection{J0030+0451}
The profile shows two $\sim$100\degr-wide components that resolve into multiple components.  The stronger one preceeds the double-peaked gamma-ray profile, whereas the weaker one lines up with the centroid of the gamma-ray profile. The PA swing of this component is relatively flat, while the strong component shows some complex PA behaviour, especially at its trailing edge which may partly overlap with the gamma-ray emission. Focusing therefore on the PA values of leading part of the strong component, ignoring the latter phases, a RVM fit results in $\alpha=66\pm3$\degr\ and $\zeta=64\pm3$\degr.

\subsubsection{J0101$-$6422}
The profile comprises two components. One is about 100\degr\ wide and is double-peaked with bridge emission. Separated by about 180\degr\ from the centre of this component is relatively simple, stronger radio feature. The first component shows linear polarisation only in its second peak, with the PA being flat over the limited range of longitudes. The strong disjoint component shows two prominent orthogonal jumps in a mostly flat PA swing. An RVM fit is not possible. The single radio peak overlaps with the gamma-ray profile, the double-peaked component does not though this statement may need to be modified (certainly for the leading part) once more S/N became available.

\subsubsection{J0154+1833}
This relatively simple two-component profile is about 70\degr\ wide. The gamma-ray profile has low s/n but it partly overlaps with the trailing edge of the radio profile, which may be the cause of the PA in this region being very flat. FAST sensitivity reveals an extra, disjoint and weak component trailing the strong component by about 120\degr. Polarisation is not detected in this component. A RVM fit to the strong component yields $\alpha=70\pm7$\degr\ and $\zeta=77\pm7$\degr.

\subsubsection{J0613$-$0200}
The profile width is 250\degr, comprising of low-level emission and multiple connected components. The gamma-ray components lag the radio components, with some overlap. The PA swing is complex. A Shapiro delay measurement \citep{2023ApJ...951L...9A} suggests a large magnetic inclination and viewing angle.

\subsubsection{J0614$-$3329}
The profile has a strong radio component and a weaker one, about 150\degr\ apart, both about 100\degr\ wide. The weaker component has a clear peak and a longer trailing tail.  The disjoint weak component is weakly polarised and the few PA values do not allow for a reliable RVM fit. The strong component has significant polarisation, but the PA is mostly very flat, again not contributing well to a reliable RVM fit. The formal fit yields $\alpha=116\pm2$\degr\ and $\zeta=118\pm2$\degr\ but is largely dominated by the attempt to model one of the presumed orthogonal jumps as a very steep swing. We do not consider this fit to be very reliable. The gamma-ray profile is double-peaked with the weaker component overlapping the trailing part of the strong radio component.

\subsubsection{J0751+1807}
High-sensitivity FAST observations \citep{whx+23} reveal that the previously known double-peaked radio component extends to far earlier longitudes, by another 70\degr. They also reveal a weak disjoint component trailing by another 70\degr\ or so. This component is too weak to discern polarization and aligns with the box-like gamma-ray profile. The strong radio component shows three clear orthogonal jumps between which the PA follows a clear RVM-like slope. A fit to these PA gives  $\alpha=114\pm5$\degr\ and $\zeta=92\pm6$\degr.

\subsubsection{J0931$-$1902}
The profile of this pulsar shows three prominent peaks, although the strongest and weakest component may be connected by bridge emission. If so, this would result in a very wide ($\sim 200$\degr) component and one narrower one ($\sim 30$\degr). The wide one is aligned with centre of the wide gamma-ray component. This would suggest that the narrow component is originating from the PC. It consists of a highly linearly polarised part, and a less polarised trailing ``shoulder''. The PA swing here is rather flat but only defined over about 15\degr\ under the strongly polarised part. The PA of the wide component is also flat with constant value over 60\degr\ when accounting for a prominent 90\degr\ jump. Those values differ from the those of the narrow component by about 40\degr. It is difficult to judge if the narrow component is part of an S-like swing expected in the RVM; the wide component aligned with the gamma-ray is clearly not RVM-like.

\subsubsection{J0955$-$6150}
The profile has a wide ($\sim225$\degr) component that shows multiple sub-components separated from a narrow ($\sim18$\degr) component of much smaller amplitude. The polarisation profile was studied in detail by \cite{psr0955}, showing that the narrow component is highly polarised with a flat PA swing. The wide component shows a low degree of linear polarisation with a PA that is mostly flat but with a U-shape feature at its centre and some non-orthogonal jump at the leading part. \citet{psr0955} fitted an RVM to the data, indicating $\alpha \sim 73$\degr\ and $\zeta \sim 77$\degr, close to the value for the orbital inclination angle of $i=83$\degr. The gamma-ray profile is double peaked and aligns perfectly with the wide component, suggesting that this dominant radio feature is originating from the LC, whereas the narrow weaker component is from the PC.

\subsubsection{J1024$-$0719}
The profile of this pulsar consists of a strong complex component with a width of nearly 200\degr, followed 150\degr\ later by an extremely weak component. The single gamma-ray component is aligned with the weak radio component but its S/N is limited and it may extend further. The PA is complex with a steep swing towards the trailing edge of the strong component and with a peculiar kink, most likely due to an orthogonal jump. 
Fitting a RVM to the strong-component does not yield a satisfactory result.

\subsubsection{J1036$-$8317}
The profile has two wide, disjoint components, and the hint of another, very weak one that needs to be confirmed. The two prominent components are separated by about 50\degr\ as measured from their edges, whereas the stronger, wider component has a low-level wing extending the width to 100\degr. The detected gamma-ray profile is positioned somewhat offset from the strongest radio component, but overlapping in large parts with the remainder of the profile. This suggests that the strong component originates from the LC. It is difficult to find an RVM that satisfactorily describes the complex PA swing. We ignore the PA values of the stronger components, this results in $\alpha=99\pm12$\degr\ and $\zeta=97\pm13$\degr\ which we view with caution. 

\subsubsection{J1125$-$5825}
The radio profile shows one dominant wide ($\sim$90\degr) feature that resolves into three components. A second disjoint double-peaked feature (that barely connects with the given S/N), itself being separated by $\sim 180$\degr. The gamma-ray profile appears wide and its main peak aligns  with the double-peaked radio feature, suggesting that this feature -- or parts of it -- result from LC emission. The dominant radio feature does not have a gamma-ray counterpart and should therefore be the PC component. Fitting an RVM to the whole profile is not possible, and hence, we applied the RVM to the strong component only. Ignoring a peculiar upturn of the PAs, possibly due to an unresolved orthogonal jump, we obtain $\alpha = 94\pm 13$\degr\ and $\zeta = 123\pm 12$\degr. 

\subsubsection{J1125$-$6014}
The profile has a dominant strong, 40\degr\ wide component, which extends at lower amplitude another 40\degr\ or so. It has a disjoint component of complex shape, which is about 210\degr\ wide. The first part of this lowly polarised wide feature shows a prominent S-like swing, whereas its trailing part shows yet another upwards PA slope. The dominant component is about 40\% linearly polarised in its first part and almost unpolarised in its trailing part, showing overall a distinct PA-swing but with ``kink'' that divides the swing into two parts with almost linear slopes, i.e.~a shallower one followed by a steep one. The gamma-ray profile has poor S/N but appears to show overlap with the wide radio component. The first part of the wide component is exactly 180\degr\ apart from the dominant radio component perhaps indicating an orthogonal geometry. It is possible to perform an RVM fit to the first PA swing of the disjoint component and the trailing part of the dominant component, resulting in an orthogonal rotator ($\alpha = 89 \pm 4$\degr\ and $\zeta = 84 \pm 4$\degr). The derived viewing angle is consistent with the measured inclination angle of $\sin i=0.978(5)$ from a Shapiro delay \citep{pptaII}.

\subsubsection{J1207$-$5050}
The profile consists of two prominent disjoint radio features, preceded by a third weaker one. The first disjoint, weaker and narrower ($\sim 20$\degr\ width) but highly linearly polarised component precedes the strong component by about 50\degr. The PA is sloped under the strongest component, but flat and with the same PA value for the other two components. The two-peaked gamma-ray profile has the strongest peak aligned with the trailing peak of the wide component and a weaker component aligned with the disjoint component. This may suggest that these two components originate from the LC, producing the same flat PA values, while the strong radio component comes from the PC, blending with radio emission from the LC. An RVM fit to the central component is however not very constraining.

\subsubsection{J1231$-$1411}
The profile shows a strong, 70\degr\ wide component with a weak, very highly polarised component trailing by about 90\degr. This component may extend further, but the S/N is too limited to tell. The gamma-ray profiles is offset from the strong component, but one of the gamma-ray peaks nearly aligns with the weak component. Accounting for a orthogonal jump, a RVM fit results in $\alpha=89\pm23$\degr\ and $\zeta=81\pm23$\degr.

\subsubsection{J1301+0833}
The radio profile extends over 200\degr\ of longitude. There are at least four components, with two small flanking components around two larger inner components. The PA swing is smooth apart from an wide `dip' due to an orthogonal jump. The gamma-ray profile consists of two peaks, the first of which clearly overlaps with the trailing two radio components. It is of limited S/N and a wider overlap cannot be excluded. For this reason, it is difficult to identify possible PC components that allow for a reliable RVM fit.

\subsubsection{J1302$-$3258}
The profile shows two simple radio components separated by approximately 150\degr. The weaker component is highly polarised and half the width ($\sim$36\degr) of the stronger component ($\sim$73\degr). The double-wing gamma-ray profile, which has a width of $\sim$210\degr, is located between the two components and partly overlaps the weaker one. It is possible to fit an RVM to the PA swing of the stronger, presumed PC component alone:
 $\alpha=101\pm14$\degr\ and $\zeta=93\pm15$\degr.

\subsubsection{J1327$-$0755}
This pulsar is difficult to interpret, as it shows emission over almost the full period. One very wide component of 235\degr\ is followed by a disjoint component about 50\degr\ later. With higher S/N, these may well be connected. The profile is only weakly polarised, not providing a sufficiently large range of well defined PA values. A convincing RVM fit is not possible. The gamma-ray profile has low S/N, but the main radio peak does not appear to have a gamma-ray counterpart.
 
\subsubsection{J1455$-$3330}
This profile has a prominent feature that is about 107\degr\ wide and a weak disjoint component, trailing by about 160\degr.
While the weak component is 100\% polarised, the strong feature is little polarised. Because of the excellent S/N, the PA is nevertheless defined over a wide range, but it is very complex. The double-winged gamma-ray profile is centred on the weak disjoint component, but with a width of about 145\degr, and it is much wider than the narrow radio component of about 25\degr\ width. In the discussed framework, the weak component would be LC emission and the strong radio component would be PC emission. Trying to fit only the central PAs, most likely associated with PC emission here, results in $\alpha =105\pm2$\degr\ and $\zeta=112\pm2$\degr, which is consistent with a Shapiro delay measurement (Pillay et al.~submitted).

\subsubsection{J1614$-$2230} 
The profile shows a prominent radio component and a wide (175\degr) low-amplitude disjoint component that resolves into several components. One of these clearly resolved component is exactly 180\degr\ apart from the main peak, which is 100\% linearly polarised, whereas the low-level feature shows little polarisation. The PAs are quite flat, and a fit of a RVM is not possible. The gamma-ray profile shows a double-wing shape with a central component that almost perfectly aligns with the weaker radio component. We note that the measured Shapiro delay yields $i=87.17\pm0.02$\degr\ (or its 180\degr-complement) \citep{2010Natur.467.1081D} suggests a similar $\zeta$ and a correspondingly large $\alpha$.

\subsubsection{J1658$-$5324 }
This source has an extremely wide profile of at least 300\degr\ width, where the baseline is difficult to define given the S/N. A prominent double peaked component is preceded by a resolved small component and a wedge-like feature. The gamma-ray profile is very wide with a smaller peaked aligned with the double-peaked radio component and extending with a large wide component to the first resolved radio component, it aligns with. The PA is defined across the whole radio profile, but it is mostly flat and with "kinks" that do not allow an RVM fit.

\subsubsection{J1730$-$2304}
This source was among the first to be identified with a disjoint component \citep{paperI}, which trails the triple-peaked wide ($\sim$150\degr) profile by about 80\degr. The high S/N reveals also another complex feature that seems to connect to the main pulse. The PA in the leading and trailing features is flat but is different by about 50\degr. The PA under the main pulse is complex. The double-wing gamma-ray profile aligns with its prominent peak with the disjoint component, extending in width to the preceding low-level radio features. This may suggest that the low-level component is LC emission, while the prominent radio peak is PC emission, as suggested already by \citet{paperII}.
Ignoring the flat PAs and fitting the presumed PC-emission alone results in $\alpha=60\pm10$\degr\ and $\zeta = 50\pm10$\degr. 

\subsubsection{J1741+1351}
In this profile a weak component precedes a simple looking strong radio component by about 100\degr. The degree of polarisation is modest, but the PA is well defined over a large range of longitude. The strongest gamma ray peak aligns with weak radio component but it does not extend all the way to the strong radio feature. A second gamma-ray peak is located at a phase trailing the strong component by about 80\degr. Ignoring PAs of the weak component,  fitting an RVM is possible, leading to $\alpha=108\pm3$\degr\ and $\zeta=121\pm2$\degr.

\subsubsection{J1744$-$1134}
In this profile, the strong radio feature is preceded by a disjoint weak, 60\degr\ wide component, leading by about 70\degr. While the strong component is nearly 100\% polarised, the weak component is not. The gamma-ray profile shows emission for nearly the full pulse period; the two radio components lie on the trailing edge of the gamma components. The PA swing is complicated and a RVM fit is only possible when ignoring some of the kinks that are clearly observed. This results in $\alpha=114\pm1$\degr\ and $\zeta=95\pm1$\degr.

\subsubsection{J1745+1017}
The profile shows a 130\degr\-wide strong radio component and two disjoint components preceding and trailing the main component. The gamma-profile is wide, covering the disjoint components in phase but being misaligned with the leading part of the strong component. The RVM fit results leads to $\alpha=88\pm6$\degr\ and $\zeta=80\pm6$\degr.

\subsubsection{J1811$-$2405}
This pulsar has a very wide profile, extending for $\sim$250\degr. In addition to a strong gaussian-like component, at least six weaker but connected components are visible. As a result there is significant overlap in phase with the gamma-ray profile which shows two distinct peaks. The PA swing consists of different, relatively flat parts that are offset to each other due to orthogonal modes. A RVM fit is not possible.

\subsubsection{J1832$-$0836}
This profile shows three  prominent features, two of which have clearly resolved components. In addition there are two extra weak components nearly halfway in between these features. The gamma-ray profiles shows a single, relative narrow component that aligns nearly perfectly with the strongest radio peak. None of the other radio components seem to have a gamma-ray counterpart. The weaker of the three strong features is 100\% linearly polarised and the PA is well defined under all of these components. One of the weak components lies exactly 180\degr\ apart from the presumed PC component. Using their PA values and ignoring the complex PA swing of the LC components results in $\alpha=100\pm2$\degr\ and $\zeta = 72\pm2$\degr.

\subsubsection{J1855$-$1436}
This rather weak profile has two components although it is possible that they are connected in a double-peak profile of about 170\degr\ width. The gamma-ray profile aligns with the first of the two components. Only the latter has sufficient linear polarisation, showing a gentle slope of about 40\degr\ over 50\degr\ longitude -- not sufficient to fit a RVM.

\subsubsection{J1921+0137}
The profile of this pulsar shows two radio components separated by 180\degr. The radio profile aligns well with the double-peaked gamma-ray profile, The stronger radio component also has a flat PA swing, while the PA slope in the other component is non-zero but relatively shallow. A meaningful RVM fit is not possible.

\subsubsection{J1939+2134}
The profile of the first millisecond pulsar discovered shows two strong radio components that are unusual as they align perfectly with two prominent gamma-ray components. The pulsar is also well known to exhibit giant radio pulsars \citep{1996ApJ...457L..81C} from similar pulse phases, suggesting that these and the strong radio components originate from the LC. Indeed, the PA swing under these components is completely flat, inconsistent with an RVM. However, high sensitivity observations reveal additional, very low-level components that may represent the PC emission. It is possible to fit a RVM to those components, resulting in $\alpha=108\pm2$\degr\ and $\zeta = 96\pm2$\degr.

\subsubsection{J1946+3417}
The profile shows radio emission over about 345\degr\ of longitude. Although the gamma-ray profile has low s/n, there may be emission of a large range of phases, where a weaker gamma-ray peak appears to line up with the second strongest radio component. The PA is mostly flat but with a downwards slope in the dominant radio component -- a fit of an RVM is not possible. A measurement of $\sin i$ \citep{aab+21} implies $\zeta\sim65$\degr.

\subsubsection{J1959+2048}
This complex profile shows a wide $\sim130$\degr-wide and two additional, apparently disjoint narrow components. The weaker component is highly polarised and has no gamma-ray counterpart. In contrast, the strong, narrow component aligns perfectly with a strong gamma-ray peak. The wide radio component exhibits a strongly polarised narrow component and a weaker, less polarised wide component. The latter has a gamma-ray counterpart that matches in both phase and shape. It is highly suggestive that the two narrow, highly linearly polarised radio features correspond to a main pulse–interpulse pair originating from the PC, whereas the weakly polarised features aligning with the gamma rays are LC components. A corresponding RVM fit yields $\alpha=104\pm2$\degr\ and $\zeta = 78\pm4$\degr.

\subsubsection{J2017+0603}
The profile shows two strong, connected components, spanning about 270\degr\ in longitude. They have similar shapes, with weaker leading and stronger trailing features. The PA values, where defined, show a flat swing, apart from four prominent orthogonal jumps. An RVM fit is not possible. The gamma-ray profile is broad (150\degr), preceding the first strong radio component.

\subsubsection{J2042+0246}
The profile shows two components of quite different amplitude but mirror-images of each other. A weak component with sharp rise and gentle decay is followed about 230\degr\ later by much stronger component with a gentle rise but sharp decline. The gamma-ray profile aligns with the components, but ``the other way around" with the stronger component showing the gentle rise on the outside rather than the inside. The PA shows some slope in the strong component, but is rather flat for the weaker one. An RVM fit may be  possible, but given the alignment between radio and gamma-ray and the peculiar shape similarities, it is possible that both components originate from the LC. In this case, the meaning of any derived angles would be questionable.

\subsubsection{J2043+1711}
The profile is complex, with many resolved components. High sensitivity show that all components are joined, forming a 290\degr-wide profile. PA values are defined over a wide range, resulting in very complex swings. It is difficult to judge which of PA values could be affected by LC emission. One attempt of a RVM fits leads to $\alpha=115\pm2$\degr\ and $\zeta=108\pm4$\degr. This can be compared to an orbital inclination angle as determined via a Shapiro-delay, yielding 82\degr\ or 98\degr \citep{2023ApJ...951L...9A}, giving some credibility to the fit.

\subsubsection{J2124$-$3358}
This is an unusually complex profile, comprising five distinguishable components that together cover almost the entire pulse period. The strongest component, which is 100\degr\ wide, may be disjoint, but the other four appear to be connected. The profile has only one highly polarised component, which is very 'peaky', and is the third of the four connected components. The first, weaker component is partially 100\% polarised. PA values are only defined under these two components, which are mostly flat. The wide gamma-ray profile closely aligns with the peak of the strongest of the four components. An RVM is possible, but its meaning is unclear. 

\subsubsection{J2150$-$0326}
The $\sim$200\degr\ separation between the strong and weak radio components may suggest a mainpulse-interpulse configuration. However, the triple-peaked gamma-ray profile has its strongest peak aligned with the weak radio component.
The strong radio component, in contrast, is located at phases where no gamma-ray emission is detected significantly. Both
components show some polarisation. The PA values of the weak, gamma-aligned component are essentially flat. The strong component 
shows peculiar non-RVM like swing. Whether this is caused by overlapping LC components is hard to say, but it prevents a meaningful RVM fit.

\subsubsection{J2215+5135}
The profile has three simple, disjoint components. Whether there is emission connecting them is not possible to say with the given S/N. The double peaked gamma-ray profile align with two of the radio peaks with the second largest radio component offset. The PA is well defined for the strongest component, which shows a quite high degree of polarisation in contrast to the other components. However, no RVM can be fitted.

\subsubsection{J2234+0944}
The profile shows three components, two very strong components plus an apparent disjoint one, preceding the first strong one by 20\degr. There are hints for an additional weak, disjoint component and likely there is emission over almost the entire pulse period. The resulting PA is also defined over almost 330\degr, but it is also very complex. The gamma-ray profile is about 150\degr\ wide, being offset to the strong radio components, but overlapping with the weak emission. Where there is overlap, the PA is relatively flat. At the other phases, we observe a lot of orthogonal jumps, contributing to the complex PA behaviour which makes a RVM fit extremely challenging.

\subsubsection{J2241$-$5236}
The radio profile shows a relatively simple component and a low-level component 180\degr\ later. In contrast, the gamma-ray profile is complex, with emission over almost the full period. The strong radio component aligns with the only gap in gamma-ray emission. The PA swing is very complex and cannot be fitted by the RVM.

\subsubsection{J2302+4442}
The profile extends over almost the full longitude range. There are three prominent peaks. The initial triangular peak aligns with the brightest gamma-ray peak. The two brightest radio peaks are joined via a central bridge and have no gamma-ray counterpart. Polarization is high in all components but the PA swing is non-RVM like. A measured Shapiro-delay suggests a viewing angle close to 90\degr \citep{2023ApJ...951L...9A}.

\subsection{Description of pulsar profiles from Table~\ref{tab:NOLC}}
\subsubsection{J0023+0923}
The profile is 90\degr\ wide, and consists of at least two connected components. The double-peaked gamma-ray component is delayed with respect to the radio components. The PA swing consists of four mostly flat parts, separated by jumps. Only the latest part shows some degree of downwards slope. An meaningful RVM fit is not possible.

\subsubsection{J0437$-$4715}
At first glance, the profile spans about 200\degr\ of longitude, consisting of multiple connected components. The single peaked gamma-ray profile lags the strong radio components. The PA swing is complex. Looking more closely and taking the PA values into account, a wide range of low-level emission is detected, spanning about 330\degr. This emission seems to overlap with the gamma-ray profile in phase, and it is unclear how it is associated with LC emission. \citet{2024ApJ...971L..18R} have measured an orbital inclination angle of 137\degr.

\subsubsection{J0653+4706}
The profile has two components separated by 165\degr. The weaker component appears to be more polarised than the strong component. An RVM fit is not possible. The PA slope of the strong components are flat after accounting for an orthogonal jump and the range of PA values defined for the weak component is too small. The gamma-ray profile has low s/n and the alignment with the radio profile is unclear.

\subsubsection{J0737$-$3039A}
The polarisation properties of this pulsar have recently been discussed by \cite{relbin}. The data confirmed an orthogonal geometry with the two strong radio components originating from two opposite polar caps. The gamma-ray profile is misaligned but overlaps partly in phase with the radio emission, at longitudes which were mostly ignored in the \cite{relbin} fit. The geometry shows an equatorward line of sight with $\alpha=79.1\pm0.1$\degr\ and $\zeta=88.4\pm0.1$\degr.

\subsubsection{J0740+6620}
For this pulsar, we use unpublished data from the Effelsberg telescope. The profile shows two strong radio peaks, both of which show a moderate degree of polarisation. The components are both about 70\degr\ wide, and both the gamma-ray peaks misalign with the radio components. The PA can be fitted with an RVM, resulting in $\alpha=67\pm2$\degr\ and $\zeta=88\pm2$\degr.

\subsubsection{J1012$-$4235}
The radio profile is narrow and relatively simple in appearance. The polarization is low and the PAs shows an apparent rapid swing through the centre of the profile but when accounting for orthogonal jumps, the PA is essentially flat, preventing a RVM fit. The gamma-ray profile consists of two components and the radio component lies between the two.

\subsubsection{J1446$-$4701}
This profile resembles that of PSR~J0437$-$4715 described earlier. A large single peak is flanked by outriders over about 110\degr\ of longitude and the PA swing steepens towards the trailing edge of the profile where there is overlap with the gamma-ray emission in phase. The boxy gamma-ray profile lags the radio components. An RVM fit is not possible.

\subsubsection{J1514$-$4946}
The profile consists of three peaks with the leading peak the brightest. The polarization fraction is high throughout. The PA swings steeply towards the edge of the profile. The gamma-ray profile is double peaked and lags the radio emission. It is possible to fit an RVM with $\alpha=82\pm10$\degr\ and $\zeta=86\pm10$\degr.

\subsubsection{J1544+4937}
The simple radio profile has only a low level of polarization and is relatively narrow. The gamma-ray profile has low S/N but its single humped peak appears to lag the radio emission. A RVM fit is not possible.

\subsubsection{J1600$-$3053}
The radio profile is relatively narrow with two major components. The PA swing is smooth after taking into account the orthogonal jump,  allowing for a RVM fit: $\alpha=152\pm12$\degr\ and $\zeta=129\pm19$\degr.  The gamma-ray profile is double-peaked and lags the radio profile.

\subsubsection{J1630$+$3734 }
The profile shows two very wide ($\sim120$\degr) components. Both components show multiple sub-components and a complex PA swing. It is still possible to fit a satisfactory RVM, if one ignores the PA of the leading part of the stronger component, where one also observes orthogonal jumps and a sense-reversal in circular polarisation. The fit suggests an orthogonal rotator, i.e.~$\alpha =90\pm1$\degr\ and $\zeta=81\pm1$\degr. The double-peaked gamma-ray profile is not aligned with the radio profile.

\subsubsection{J1640+2224}
This profile is similar to that of J1600$-$3053 above. An apparent narrow radio profile with several components precedes the double-peaked gamma-ray profile. Low level emission is detected for a about 140\degr. The PA swing is almost flat but shows peculiar jumps and an upward turn at its trailing part. A RVM fit is not possible.

\subsubsection{J1713+0747}
In some ways, this profile shows similarities to PSR J0437$-$4715. At first glance, the profile is restricted to a width of about 120\degr. However, high sensitivity observations reveal emission over more than 330\degr. The single-hump gamma-ray profile is wide and offset from the radio peak, but lines up with this extended low-level emission. The resulting PA swing comprises several gently curved parts that jump due to orthogonal modes but are overall inconsistent with an RVM. Since it is difficult to separate possible LC from PC components, a reliable RVM fit is impossible. 

\subsubsection{J1732$-$5049}
The profile covers some 240\degr\ of longitude with a slowing rising ramp followed by the major peak and two subsequent weaker components. The PA swing is very shallow before steepening at the profile edge. The pulsar was among a group of MSPs identified by \cite{relbin} as beloning to a class of flat-PA pulsars. The single humped gamma-ray profile fits perfectly into the off-pulse radio profile. A RVM fit is not possible.

\subsubsection{J1857$+$0943}
The profile shows two strong radio components separated by about 180\degr. Both components are about 90\degr\ wide. The degree of polarisation is low in both components. However, the PA is well defined over a wide range of phases. Knowing the orbital inclination angle could constrain the RVM solution effectively, but the gamma-ray profile has a low signal-to-noise ratio. This prevents us from determining whether the partly very complex PA swing is affected by LC components.

\subsubsection{J1903$-$7051}
The profile is 150\degr\ wide and consists of a small leading and trailing component with the radio peak situated in between. The PA swing is smooth, with a steepening towards the edge of the profile. The single-peaked gamma-ray profile lags the radio components. An RVM fit yields $\alpha=105\pm2$\degr\ and $\zeta=86\pm2$\degr.

\subsubsection{J1909$-$3744}
In addition to the strong radio peak, there is an interpulse feature about 200\degr\ apart. There is a very weak disjoint component trailing the main component by about 30\degr. The PA is only well defined in the strong component but with some values available for the interpulse. The measured Shapiro delay implies a viewing angle of close to $\zeta\sim86$\degr\ or $\zeta\sim94$\degr\ \citep{pptaII}. Using this constraint, the solution $\alpha=81\pm2$\degr\ and $\zeta = 85\pm2$\degr\ is preferred over the complement. The gamma-ray profile is extremely weak and nothing can be discerned about the alignment between gamma-rays and radio.

\subsubsection{J1946$-$5403}
The profile is narrow with little linear polarization. There are not enough PA points to determine a trend or an RVM solution. The double-peaked wide gamma-ray profile perfectly misaligns with the radio profile.

\subsubsection{J2039$-$3616}
The radio profile consists of three peaks with the trailing component being the strongest. The polarization fraction is high throughout and the PA swing is flat and broken by an orthogonal jump. A RVM fit is not possible. The single humped gamma-ray profile lags the radio.

\subsubsection{J2317+1439}
High-sensitivity observations reveal emission over 250\degr\ of longitude. The prominent strong feature alone extends over 100\degr\ with two prominent peaks followed by a weak trailing component. The polarization is moderate and the PA swing is complex. The gamma-ray profile has low S/N and little can be determined about the overlap with the radio emission.

\subsection{Description of pulsar profiles from Table~\ref{tab:DNO}}
\subsubsection{J0406+3039}
The profile shows a strong, very wide ($\sim$180\degr) component and a disjoint component trailing by about 60\degr, itself about 70\degr\ wide. The later parts of the strong component are highly linearly polarized. Overall, the PA swing is quite complex, especially for the leading, much less polarised part of the strong component. Despite a ``kink'' in the PA swing, one can attempt to fit an RVM connecting all the PAs. However, many solutions assign the fiducial plane to a location outside the strong component. The corresponding fiducial plane 180\degr\ later misses both components altogether. We therefore do not believe this fit to be reliable.

\subsubsection{J0645+5158}
This pulsar exhibits a prominent narrow component, followed by a weaker component approximately 135\degr\ later. This separation suggests that the weaker component is not an interpulse. PA values can be measured for both components. The swing under the strong component is complex and inconsistent with the RVM. The PA swing of the weaker component is rather flat and, overall, a meaningful RVM fit is not possible.

\subsubsection{J0824+0028}
The profile has three prominent components, two of which are clearly connected. The third component appears disjoint and precedes the pair, showing the same flat PA values as the other components. This provides a gentle, small slope of PA across the profile. This suggests that, with higher sensitivity, all components would be connected, forming a profile wider than 250\degr. Fitting the PA suggests an aligned rotator with $\alpha=176\pm2$\degr\ and $\zeta=169\pm5$\degr.

\subsubsection{J0900$-$3144}
The profile shows a $\sim150-180$\degr\ wide strong feature, with a dominant relatively simple component to lead and a double-peaked component of half the strength following. There is also a very weak component separated by about 180\degr\ from the phase that connects the two strong radio component. The PA values of the strong feature are flat. The weak component too weak to have significantly measured PA. Fitting the PA suggests an aligned rotator with $\alpha=5\pm2$\degr\ and $\zeta=13\pm6$\degr.

\subsubsection{J0921$-$5202}
The strong and weak components are likely connected by low-level emission, resulting in a width of approximately 180\degr. As the components appear to be joined, the downward PA slope in the weak component and the upward slope in the strong component are unusual and not RVM-like. In any case, there are insufficiently defined PA values to fit an RVM.

\subsubsection{J1216$-$6410}
The profile shows an about 50\degr-wide double-peaked component, separated from a broad ($\sim170$\degr) component by about 50\degr. No gamma-rays have been detected. The PA swing of the broad component is relatively flat. In contrast, the PA swing of the narrow component is steep. Fitting an RVM only to the strong component, and ignoring a downward part of the PA slope at the trailing edge, one finds an inner line of sight: $\alpha=61\pm17$\degr\ and $\zeta=57\pm16$\degr.

\subsubsection{J1529$-$3828}
A 90\degr-wide, featureless component is separated from a disjoint, much weaker component by 180\degr. It is difficult to tell whether both of these originate from the PC. As expected given the low gamma-ray flux, no high-energy emission has been detected. As the PA swings in both components are very flat, no RVM fit was possible.

\subsubsection{J1547$-$5709}
The double-peaked strong component is about 100\degr\ wide. There is a weak leading component which appears to be disjoint, but with the available S/N it is difficult to assess if there is a low level of emission connecting the features. There is a gentle slope in the PA values of the strong feature, but only few PA values are available for the weak component. Still, it is possible to fit an RVM to all PA values, assuming that they are all PC components, which is far from certain giving the non-detection of gamma-rays. The RVM fit yields  $\alpha=42\pm8$\degr\ and $\zeta=75\pm157$\degr.

\subsubsection{J1603$-$7202}
The 80\degr\ wide main component sits atop low-level emission which stretches the width to 180\degr. An apparent disjoint weak component is located about 100\degr\ after the strong component. Parts of the PA swing look RVM-like, but most of the measured PAs are flat. Using those PA which show an RVM-like swing, a corresponding fit yields $\alpha=30\pm12$\degr\ and $\zeta=27\pm12$\degr. The viewing angle is consistent with the orbital inclination angle of $i=31\pm3$\degr\ determined from scintillation measurements \citep{2022ApJ...933...16W}. 

\subsubsection{J1618$-$3921}
In addition to the about 80\degr-wide strong, double-peaked component, one can observe a weak, apparent disjoint component preceding the strong component by about 70\degr. The given S/N does not reveal any connecting emission. The PA values are rather flat and are not well described by a RVM. 

\subsubsection{J1709+2313}
A strong radio feature with a resolved component on its leading edge is followed, approximately 100\degr\ later, by a highly polarised component of approximately half the strength. Although the current signal-to-noise ratio is insufficient, the two features may be connected by low-level emission, which would make the profile approximately 250\degr\ wide. If the components are instead disjoint and originate from different locations, perhaps the highly polarised component originates near the LC. Without this knowledge, a reliable RVM fit is not possible.

\subsubsection{J1731$-$1847}
A strong, modestly polarised component is followed, about 140\degr\ later, by a disjoint, little-polarised component. Despite this source having a 4FGL counterpart, no pulsed gamma-ray emission has been detected. The PA values of the strong component are not characteristic of the RVM and a fit is not possible.

\subsubsection{J1804$-$2717}
The peculiar broad component has two prominent peaks of equal amplitude and exhibits a high degree of linear and circular polarisation, particularly in the trailing region. About 180\degr\ separated from the centre of the broad component, a weak, disjoint component can be detected. The high polarisation of the weak component leads to the measurement of PA values, but the narrow range of longitudes does not permit the detection of a slope. The strong component exhibits a modest slope in the PA values, accompanied by a 'kink' associated with the first peak, a reversal in the sense of circular polarisation, and potentially an orthogonal jump. It is possible to fit all PA values with a single RVM, leading to an aligned geometry,  $\alpha=160\pm7$\degr\ and $\zeta=144\pm12$\degr, but it is not clear how meaningful the application of the RVM is in this case.

\subsubsection{J1826$-$2415}
There are two components, one of which is much weaker than the other, each with a width of about 100\degr. They are separated by approximately 70\degr, with their respective centroids around 180° apart. Both components are essentially unpolarised, so an RVM fit is not possible.

\subsubsection{J1906+0055}
With limited S/N, the profile has a 100\degr-wide component, with a peak and a 'shoulder'. A similarly shaped, somewhat weaker component is visible 90\degr\ earlier. The profile shows no detected polarisation, preventing further discussion.

\subsubsection{J1918$-$0642}
A triangular-shaped strong component with a central, narrow peak is separated from a weaker component by about 180\degr. It is strong enough to reveal significant PA values of a complex swing.  The PA swing of the strong components is even more complex with a number of wiggles, making a meaningful RVM fit impossible.

\subsubsection{J1923+2515}
The profile shows two disjoint components separated by some 200 \degr. The strong component has two bright peaks preceded by a weak pre-cursor like sub-component. The weaker second component is itself double peaked. The components are moderately polarised with the PA defined over a wide range of longitudes. A RVM fit is only possible, when taking into account only the PAs at the leading part of the weak component and the trailing part of the strong component. This results in an orthogonal geometry. i.e.~$\alpha=99\pm2$\degr\ and $\zeta=94\pm2$\degr. However, significant parts of the PA swing need to be ignored, casting doubt on the meaning of this fit.

\subsubsection{J1935+1726}
A broad component with two distinct peaks, spanning approximately 120\degr, is followed by a weak, disjoint component located around 180\degr\ from the centre of the broad component. The broad component shows three peaks of linear intensity, each with a flat PA swing. The weak component is unpolarised. An RVM fit is not possible.

\subsubsection{J2053+4650}
This profile resembles a main pulse-interpulse configuration, with two radio features that are 180\degr\ apart. Both features are approximately 50\degr\ wide. Their shape is simple, although the stronger components exhibit two distinct peaks. The degree of polarisation is modest, but the high S/N allows PAs to be detected over a wide range. An RVM fit is possible that connects the PAs of both components, but the reliability of the fit is questionable.

\subsubsection{J2145$-$0750}
In addition to the well-known precursor, there is an apparently disjoint, wide and very weak component that follows the wide main pulse. FAST sensitivity reveals that this component may be connected to the main pulse, though this is unclear. The pre-cursor is the only component that is highly polarised; the others are much more weakly polarised. However, the high S/N enables us to observe a complex PA swing over a wide range of longitude. It is not possible to obtain an RVM fit without ignoring most of the PA values.

\subsubsection{J2236$-$5527}
Three apparently disjoint components are visible. The strongest is about 50\degr\ wide shows three peaks and polarisation in its first half with a distinct PA slope and two orthogonal jumps. The other components are much weaker, separated from the strong component and from each other by about 45\degr, respectively. In the wider one of these, some polarisation and flat PA swings are visible. Fitting an RVM only to the strong component results in $\alpha=97\pm26$\degr\ and $\zeta=82\pm25$\degr. 

\subsubsection{J2322+2057}
The separation of two strong radio components by 180\degr\ suggests an mainpulse-interpulse case. Despite a weak degree of polarisation in both components, the strength of the pulsar allows to meausre PA values over a wide range. The somewhat weaker
component's PA swing looks RVM-like. In contrast, the PA values of the stronger components are complex, with both upward and downward slopes. Fitting an RVM is impossible without further information, such as gamma-ray detection.

\subsection{Pulsars in Table~\ref{tab:CLUM} }
This section discusses pulsars with contiguous profiles, which have a gamma-ray flux above the Fermi threshold, yet no such emission has been detected. We surmise this is due to unfavourable viewing geometry.

\subsubsection{J0509+0856}
The profile is complex with several broad components spanning in total about 330\degr\ in longitude. Apart from distinct PA slope associated with a weak, little polarised component, the remaining PA swing is amazingly flat over 300\degr. Such PA swing can be explained by an aligned geometry ($\alpha=8\pm2$\degr, $\zeta=56\pm19$\degr) but it does not describe some wiggles well. 

\label{dis0509}

\subsubsection{J0557+1550}
The profile is simple, only 38\degr\ wide. Polarisation is detected, especially in the trailing part. But, overall, the range of longitudes is too small for a meaningful RVM fit.

\subsubsection{J0711$-$6830}
The emission covers about 290\degr\ in longitude, divided into three prominent joint components. The PA is mostly flat after accounting for orthogonal jumps, leading to an aligned geometry as suggested by a RVM fit: $\alpha=6\pm3$\degr\ and $\zeta=30\pm15$\degr.

\subsubsection{J0732+2314}
This is another profile stretching over about 300\degr\ in longitude and displaying a large number (about 10) recognizable components. In contrast, the PA swing is smooth with some prominent orthogonal jumps and some  swing at the leading part of the profile. Accounting for swing as a resolved orthogonal jump, an RVM fit points to an aligned geometry:  $\alpha=29\pm2$\degr\ and $\zeta=30\pm3$\degr.

\subsubsection{J1012+5307}
The high sensitivity observations of FAST show that all components in the profile are connected. The profile is highly polarised in all its components. The resulting PA swing is very smooth, comprising of a wide flat part and one with a gentle downward slope, both separated by an orthogonal jump. It is only possible to fit a RVM by ignoring certain parts, making any fit unreliable.

\subsubsection{J1017$-$7156}
The profile is simple, gaussian-like with a weaker trailing shoulder component. With some weak emission, the total width is about 80\degr. The PA swing is complex with an u-shaped `kink'.

\subsubsection{J1045$-$4509}
The pulsar has two components, one much stronger than the other. Their centroids are located approximately 180\degr\ apart. However, the components are wide, at 95\degr\ and 110\degr\ respectively. Additionally, measurable PA values exist between the weak and strong components, suggesting that this is not an orthogonal rotator. The profile width is therefore at least 300\degr. The PA swing flat in the first component, but the strong component's PA shows distinct swings. A meaningful RVM fit is not possible for the whole range but concentrating on the strong component yields $\alpha=147\pm7$\degr\ and $\zeta=132\pm10$\degr.

\subsubsection{J1300+1240}
The profile has a simple, slightly asymmetric shape and a width of about 90\degr. The leading part is highly polarised. The resulting PA is initially relatively flat and then shows a steep downturn. A RVM fit yields $\alpha=32\pm13$\degr\ and $\zeta=36\pm14$\degr. We note that the three planets discovered around this pulsar have orbital inclination angles in the range from 47\degr\ to 53\degr.

\subsubsection{J1421$-$4409}
This peculiar profile seems to be composed of four, maybe five, Gaussian components, where the trailing one dominates. The measured PA values reveal further low-level emission for most of the period, suggesting a width of more than 300\degr\ and much wider than the 210\degr\ of the strong feature. The PA swing of the low-level emission is flat, orthogonal to the flat PAs of the dominant trailing component. The remaining parts are also  mostly flat, jumping by 90\degr\ four to five times. There is some slope at about the centre of the strong feature. Attempting to fit an RVM results is very difficult due to these jumps.

\subsubsection{J1453+1902}
FAST sensitivity reveals emission over the whole profile. The profile is dominated by a strong central component and two weaker ones. The PA swing is relatively flat when accounting for orthogonal jumps but some significant swing is prominent under the strong component. A RVM fit yields $\alpha=35\pm4$\degr\ and $\zeta=50\pm4$\degr.

\subsubsection{J1537$-$5312}
This double peaked profile has significant `bridge-emission', leading to a width of about 140\degr. Each of the peaks has some polarisation, but the defined PA values are quite flat, separated by orthogonal modes. An RVM fit yields $\alpha=44\pm18$\degr\ and $\zeta=51\pm16$\degr.

\subsubsection{J1552$-$4937}
The profile has a seemingly simple, Gaussian-like profile showing slight asymmetry. However, closer inspection reveals three components of increasing amplitude. The width is approximately 75°. The degree of polarisation is low. Where the PA is defined, it is flat with some gentle slopes interrupted by orthogonal jumps. An RVM with nearly aligned geometry can describe the observed PA, but due to the limited PA range, this is quite uncertain:
 $\alpha=32\pm11$\degr\ and $\zeta=25\pm10$\degr.

\subsubsection{J1643$-$1224}
The symmetry of this simple profile is broken by a weak but wide leading component. The total profile width is about 235\degr. The PA swing shows a slope in the trailing part of the component, and is relatively flat for the rest.  Fitting a RVM one finds $\alpha=167\pm5$\degr\ and $\zeta=170\pm4$\degr.

\subsubsection{J1653$-$2054}
The broad complex profile has a width of 260\degr. PA values can be measured for most profile phases. While most PA values are flat, near the central strong component there is initially a gentle downward slope, followed by an upward slope. An RVM fit yields  $\alpha=169\pm6$\degr\ and $\zeta=164\pm10$\degr.

\subsubsection{J1708$-$3506}
The broad profile shows one prominent peak and a weaker wide component. Some low-level emission lets the profile extend to 195\degr\ width. The PA is flat initially, shows an orthogonal jump before it slopes downwards. All PAs can be described with a single RVM, yielding $\alpha=31\pm5$\degr\ and $\zeta=55\pm5$\degr.

\subsubsection{J1710+4923}
A triple profile with a dominant central component is followed by an additional weaker component, breaking the apparent symmetry. In total, the emission spans essentially the entire pulse period. The PA is extremely complex, with a steep slope, wrap-arounds and orthogonal jumps. A RVM fit suggests a geometry very close to alignment: $\alpha=4\pm2$\degr\ and $\zeta=3\pm2$\degr.

\subsubsection{J1719$-$1438}
The profile is triangular with a very prominent peak, extending over 110\degr, and exhibits modest polarisation. The PA swing is complex, but this appears to be mostly the result of a resolved orthogonal mode, as revealed by a drop in linearly polarised intensity. Taking this into account, the RVM yields an aligned geometry given by $\alpha=168\pm6$\degr\ and $\zeta=167\pm7$\degr.

\subsubsection{J1721$-$2457}
The broad, featureless, triangular shape exhibits polarisation in its leading part. Here, the PA rises linearly and then turns over at the edge of the defined PAs. This can be described by an RVM with $\alpha=44\pm5$\degr\ and $\zeta=32\pm6$\degr.

\subsubsection{J1738+0333}
The double-peaked profile has a width of approximately 130\degr. The first, weaker component is more highly polarised than the second. The PA is defined over very a wide range and remains mostly flat until the strong component, after which it clearly turns upwards. This is where an RVM fit puts the centroid with $\alpha=159\pm8$\degr\ and $\zeta=150\pm6$\degr.

\subsubsection{J1751$-$2857}
This relatively simple profile shows a strong leading component and a much weaker shoulder component. The PA can be measured over almost the entire width of the profile, which is 100\degr. However, the flatness of the PA and the limited range of usable longitude make achieving a reliable fit difficult. It yields  $\alpha=28\pm12$\degr\ and $\zeta=27\pm12$\degr.

\subsubsection{J1801$-$1417}
The profile looks similar to that of PSR J1751--2857 but it is significantly wider, i.e.~about 150\degr. Also here, the PA is mostly flat, but the longer PA range appears to make it somewhat more reliable. An RVM fit yields 
$\alpha=169\pm6$\degr\ and $\zeta=165\pm8$\degr.

\subsubsection{J1802$-$2124}
The profile is narrow, about 30\degr\ wide. There is a steep rise in PA swing with noticeable jumps before it becomes flat. A reliable RVM fit is difficult, but one solution yields
$\alpha=146\pm15$\degr\ and $\zeta=144\pm16$\degr.

\subsubsection{J1813$-$2621}
The profile is about 95\degr\ wide and is relatively simple, showing one resolved and two overlapping components of almost equal amplitude. The degree of polarisation is quite strong, but the PA is mostly flat, with a peculiar change in slope at the trailing edge. This and the limited range of PA longitudes make the RVM relatively uncertain.

\subsubsection{J1828+0625}
The profile shows a wide base that has a triangular, three-component shape superimposed. The resulting width is about 145\degr. Fitting an RVM suggests an aligned geometry, i.e.~$\alpha=15\pm5$\degr\ and $\zeta=5\pm2$\degr.

\subsubsection{J1844+0115}
This very simple profile has a width of about 60\degr\ and does not show sufficient polarisation to conduct a RVM fit.

\subsubsection{J1850+0124}
The 100\degr-wide, double-peaked profile appears to have a weak trailing 'shoulder' component. The latter part is highly polarised, but the PA swing is very flat and not defined over a sufficiently large range of longitude. No RVM fit is possible.

\subsubsection{J1853+1303}
A strong, 80\degr-wide, triple-peaked feature is connected by low-level emission to a much weaker feature that is also triple-component and 55\degr\ wide. Together, they cover approximately 220\degr\ in longitude. The PA swing is unusual. Three peaks can be seen in the PA swing under the strong components, whereas the weaker component's PA swing is mostly flat. The peculiar peaks seem to arise from orthogonal modes. Accounting for them yields a RVM fit with  $\alpha=29\pm3$\degr\ and $\zeta=46\pm5$\degr.

\subsubsection{J1903+0327}
This simple profile appears to be heavily scatter-broadened. The PA is relatively flat, also as a result of the scattering. The geometry and an intrinsic width cannot be determined, but we note that a Shapiro-delay suggests an orbital inclination angle of either 77\degr\ or 103\degr\ \citep{2011MNRAS.412.2763F}.

\subsubsection{J1905+0400}
The profile is about 200\degr\ wide and very simple. The detected polarisation leads to a very flat PA swing and one jump at the trailing edge. The range of available PA values is too small to make a RVM fit.

\subsubsection{J1910+1256}
The profile is simple, with a width of approximately 40\degr. It exhibits linear polarisation and a distinct sense reversal in its circular polarisation. The PA swing is very steep, with flat sections at the beginning and end. It is unclear whether the steep swing is due to a central cut or an orthogonal mode. Fitting an RVM nevertheless leads to $\alpha=172\pm6$\degr\ and $\zeta=171\pm6$\degr.

\subsubsection{J1911$-$1114}
The profile is 160\degr\ wide and has a weak component that is 100\% linearly polarised and exhibits a flat PA swing. The trailing two components are much less polarised. The stronger of these two components displays an upward PA swing, while the weaker component shows a downward swing and circular polarisation as strong as the linear component. The result is a complex PA swing. We focus on the stronger component as this describes the overall PA swing best. A RVM fit yields
 $\alpha=144\pm2$\degr\ and $\zeta=141\pm2$\degr.

\subsubsection{J1911+1347}
This 240\degr-wide pulsar profile resembles that of PSRs~J0437$-$4715 and J1446$-$4701. The PA is defined over a range, showing a mostly flat behaviour, only interrupted by orthogonal modes. A geometry of  $\alpha=36\pm2$\degr\ and $\zeta=55\pm10$\degr\ describes the PA swing relatively well.

\subsubsection{J1928+1245}
The profile is simple, very slightly asymmetric and has a width of about 60\degr. There is very little polarisation, and the PAs that are defined are flat. The range of longitudes for which PA values are available is too limited for an RVM fit.

\subsubsection{J1929+0132}
The shape is similar to PSR J1017$-$7156, but this profile is somewhat wider, spanning more than 100\degr. The degree of polarisation is very low, as is the number of defined PAs. Overall, the PA swing appears flat, but it is not possible to fit an RVM.

\subsubsection{J1933$-$6211}
The profile of this pulsar is almost a mirror image of that of PSR J1828+0625. Its width is just above 100\degr. There are two adjacent peaks in both linear and circular polarisation, but the PA is defined over a wider range. The resulting PA swing is complex. Ignoring a swing presumably caused by a resolved orthogonal mode and a peculiar update at the trailing edge, an RVM fit yields  $\alpha=30\pm5$\degr\ and $\zeta=41\pm7$\degr.

\subsubsection{J1937+1658}
The first sharp component of the profile is followed by a double-peaked component, producing a profile with a width of about 160\degr. A high degree of polarisation is only visible in the trailing part, meaning the range of PAs is too narrow for an RVM fit.

\subsubsection{J1944+0907}
The profile itself is somewhat similar to that of PSR J2145$-$0750, although the central component is stronger. It is wide, spanning 335\degr. Although the degree of polarisation is low, the strength of the source means that the PA is defined across the entire profile. The resulting PA swing is complex, comprising steep slopes interspersed with orthogonal modes. Taking the latter into account, it is possible to fit an RVM yielding  $\alpha=17\pm6$\degr\ and $\zeta=9\pm3$\degr.

\subsubsection{J2010$-$1323}
Apart from a sharp peak on the leading edge, the profile is simple. Its width is 47\degr. The degree of polarisation is modest, but the PA is well defined and shows a steep upward swing at the trailing edge with some kinks' that are most likely caused by orthogonal modes. A RVM fits yields $\alpha=7\pm4$\degr\ and $\zeta=6\pm4$\degr.
 
\subsubsection{J2019+2425}
The profile shows two disjoint components, symmetrically placed around a strong, nearly featureless component. The components appear to be connected with low-level emission, making a very wide profile, covering more than 300\degr\ of longitude. The degree of polarisation is modest and relatively high only in the trailing weak component. The resulting PA values are defined over almost the entire profile, and apart from a severe ``kink'' in the strong component (clearly associated with an orthogonal jump), a ``kink'' in the opposite direction in the first weak component followed by a gentle slope, the PA is remarkable flat over most of the profile. A solution that describes the PA swing is given by 
$\alpha=3\pm2$\degr\ and $\zeta=4\pm3$\degr.

\subsubsection{J2022+2534}
The profile has a leading rising component with a prominent peak, followed by a strong component. The visible PA swing rises and transitions (ignoring the part jumped by 90\degr) into a very flat part. A RVM fit is difficult but the suggested geometry is very close to alignment, i.e. $\alpha=2\pm2$\degr\ and $\zeta=3\pm2$\degr.

\subsubsection{J2033+1734}
This simple profile has a long, weak tail that extends the width to almost 140\degr. The PA is mostly flat but has an upward trend after an orthogonal jump. An RVM fit yields  $\alpha=146\pm5$\degr\ and $\zeta=140\pm5$\degr.

\subsubsection{J2129-5721}
The profile has a width of about 180\degr. It consists of a strong double-peaked feature and a weak trailing component. The PA swing has a downwards slope and shows orthogonal jumps. Accounting for those, an RVM fit yields  $\alpha=29\pm3$\degr\ and $\zeta=47\pm5$\degr.

\subsubsection{J2144$-$5237}
The profile comprises four components, one of which has a prominent shoulder. Apart from this shoulder, all components demonstrate significant polarisation. The corresponding PA swing is mostly flat, especially under the strong components. Ignoring these PA values, it is possible to produce a fit, 
$\alpha=27\pm11$\degr\ and $\zeta=26\pm11$\degr, but this should be viewed with caution. 

\subsubsection{J2229+2643}
The profile is about 90\degr wide, exhibiting a strong peak and a preceding, unresolved component. A modest degree of polarisation allows to detect the PA over the full with, showing a mostly flat PA swing with significant kinks and a gentle slope. Ignoring the kinks, an RVM fit suggests an aligned geometry: $\alpha=5\pm3$\degr\ and $\zeta=10\pm7$\degr.

\subsubsection{J2234+0611}
The apparently simple, narrow profile appears to have a width of only 30°. However, there is low-level emission which expands the profile to approximately 250\degr. The PA starts off flat, but then shows a gentle, steady downward slope. The RVM fit suggests an aligned geometry: $\alpha=6\pm3$\degr\ and $\zeta=13\pm8$\degr.

\bsp	
\label{lastpage}
\end{document}